\documentclass[%
prl,
reprint,
superscriptaddress,
preprintnumbers,
nofootinbib,
amsmath,amssymb,
aps,
]{revtex4-1}
\pdfoutput=1
\usepackage{hyperref}
\usepackage{graphicx}% Include figure files
\usepackage{subfigure}

\usepackage[textwidth=1.25cm,textsize=footnotesize,
]{todonotes}

\usepackage[utf8]{inputenc}
\usepackage{url}

\usepackage{color}
\usepackage[normalem]{ulem}

\newcommand{\be}{\begin{equation}}
\newcommand{\ee}{\end{equation}}
\newcommand{\ba}{\begin{eqnarray}}
\newcommand{\ea}{\end{eqnarray}}

\def\jcap{JCAP}

\def\nustar{\emph{NuSTAR}}
\def\ergscm{erg~s$^{-1}$ cm$^{-2}$ deg$^{-2}$}

\def\acxb{\textit{a}CXB}

\begin{document}
\title{Strong limits on keV-scale galactic sterile neutrino dark matter with stray light \\ from NuSTAR after 11 years of operation}

\author{R.\,A.\,Krivonos}
\email[]{krivonos@cosmos.ru}
\affiliation{%
  Space Research Institute of the Russian Academy of Sciences, Moscow 117997, Russia}
\affiliation{%
  Institute for Nuclear Research of the Russian Academy of Sciences, Moscow 117312, Russia}
  
\author{V.\,V.\,Barinov}
\email[]{barinov.vvl@gmail.com}
% \affiliation{%
  % Physics Department, M. V. Lomonosov Moscow State University, Leninskie Gory, Moscow 119991, Russia}
\affiliation{%
  Institute for Nuclear Research of the Russian Academy of Sciences, Moscow 117312, Russia}

\author{A.\,A.\,Mukhin}
\email[]{amukhin@cosmos.ru}
\affiliation{%
  Space Research Institute of the Russian Academy of Sciences, Moscow 117997, Russia}
\affiliation{%
  Institute for Nuclear Research of the Russian Academy of Sciences, Moscow 117312, Russia}
\affiliation{%
  Moscow Institute of Physics and Technology, Dolgoprudny 141700, Russia}

\author{D.\,S.\,Gorbunov}
\email[]{gorby@ms2.inr.ac.ru}
\affiliation{%
  Institute for Nuclear Research of the Russian Academy of Sciences, Moscow 117312, Russia}
\affiliation{%
  Moscow Institute of Physics and Technology, Dolgoprudny 141700, Russia}

%\author{Friends}
%\affiliation{%
%  Space Research Institute of the Russian Academy of Sciences, Moscow 117997, Russia}

\bigskip
%%\date{\today}
\preprint{INR-TH-2024-012}

%Dark matter sterile neutrinos radiatively decay in the Milky Way, which can be tested with searches for almost monochromatic photons in the x-ray cosmic spectrum. 
\begin{abstract}
Using tremendous photon statistics gained with the stray light aperture of the \nustar\ telescope over 11 years of operation, we set strong limits on the emission of close to monochromatic photons from the radiative decays of putative dark matter sterile neutrinos in the Milky Way. In the energy range of 3$-$20~keV covered by the \nustar, the obtained limits reach the bottom edge of theoretical predictions of realistic models where sterile neutrinos are produced in the early Universe. Only a small region is left to explore, if the sterile neutrinos form the entire dark matter component.
\end{abstract}

%\begin{keyword}
%\end{keyword}

 %!FIX PACS!
\maketitle
%%%%%%
%%%%%%
%%%%%%

%{\bf Title selection (vote or suggest yours!):}
%\begin{enumerate}
%    \item All-sky limits on sterile neutrino galactic dark matter obtained with NuSTAR stray-light aperture background after 11 years of operation
%    \item All-sky limits on sterile neutrino galactic dark matter with stray light from NuSTAR based on 11 years of operations
%    \item Tight/Strong constraints on galactic sterile neutrino dark matter with stray light from NuSTAR based on 11 years of operations (Vlad: My vote)
%\end{enumerate}

{\it 1. Introduction.} 
A steadily growing amount of astrophysical and cosmological data persistently point at the lack of gravitational potentials in galaxies and galaxy clusters in the present Universe and at various spatial scales in the past, starting from an epoch when the primordial plasma temperature well exceeded 1\,eV. These phenomena are widely known as {\it dark matter phenomena} and have no viable explanation within the Standard Model of particle physics (SM) and General Relativity. Naturally, they can be explained by introducing additional species to the particle physics, and there are many candidates with different physical motivations, different production mechanisms operating in the early Universe and different strategies to trace their presence. 

In this letter we consider {\it sterile neutrinos} $\nu_s$ (for brief reviews see \cite{Volkas:2001zb,Gorbunov:2014efa}) of keV-scale mass $m_s$ as the viable dark matter candidate, which are singlet with respect to the SM gauge group fermions yielding masses to active neutrinos via mixing 
%within the see saw type I mechanism\,
\cite{Schechter:1980gr}. The same mixing is responsible for the sterile neutrino production in the early Universe via oscillations of active neutrinos in the lepton-asymmetric primordial plasma at temperatures about 100\,MeV\,\cite{Shi:1998km}. The same mixing induces the radiative decay of the dark matter sterile neutrino,    
\begin{equation}
  \label{decays}
\nu_{s} \rightarrow \nu_{e, \mu, \tau} + \gamma\,,
\end{equation}
which rate is\,\cite{PhysRevD.25.766,Barger:1995ty}
\begin{align}\label{eq:neutrino_width}
\nonumber\Gamma_{\gamma} &= \frac{9}{1024} \frac{\alpha}{\pi^4} G_F^2 m_s^5\sin^22\theta\\
&= 1.36 \times 10^{-22}\left(\frac{m_s}{1 \text{keV}}\right)^5 \sin^22\theta\hspace{0.25cm} \text{s}^{-1},
\end{align}
with mixing angle $\theta$, Fermi constant $G_F$ and fine-structure constant $\alpha$. The decay \eqref{decays} gives a peak-like signature at energy $E_\gamma\approx m_s/2$ in the Galactic $X$-ray spectrum\,\cite{Abazajian:2001vt}. The expected signal from a given direction is proportional to the decay rate \eqref{eq:neutrino_width} and the dark matter number density integrated along the line of sight. The entire signal spectrum is obtained by integration over the whole sky weighted with the telescope exposure.

%leptogenesis and/or constructing the  main dark matter component from sterile neutrinos, see e.g.\,\cite{Boyarsky:2009ix,Drewes:2013gca}. 

%must be light, with mass $m_s$ in keV range, see e.g.\,\cite{Abazajian:2017tcc,Dasgupta:2021ies}. However, it is still unstable, and can decay radiatively, i.e. into an active neutrino and photon, 

\begin{figure}[!b]
	\centering
	\includegraphics[width=\columnwidth]{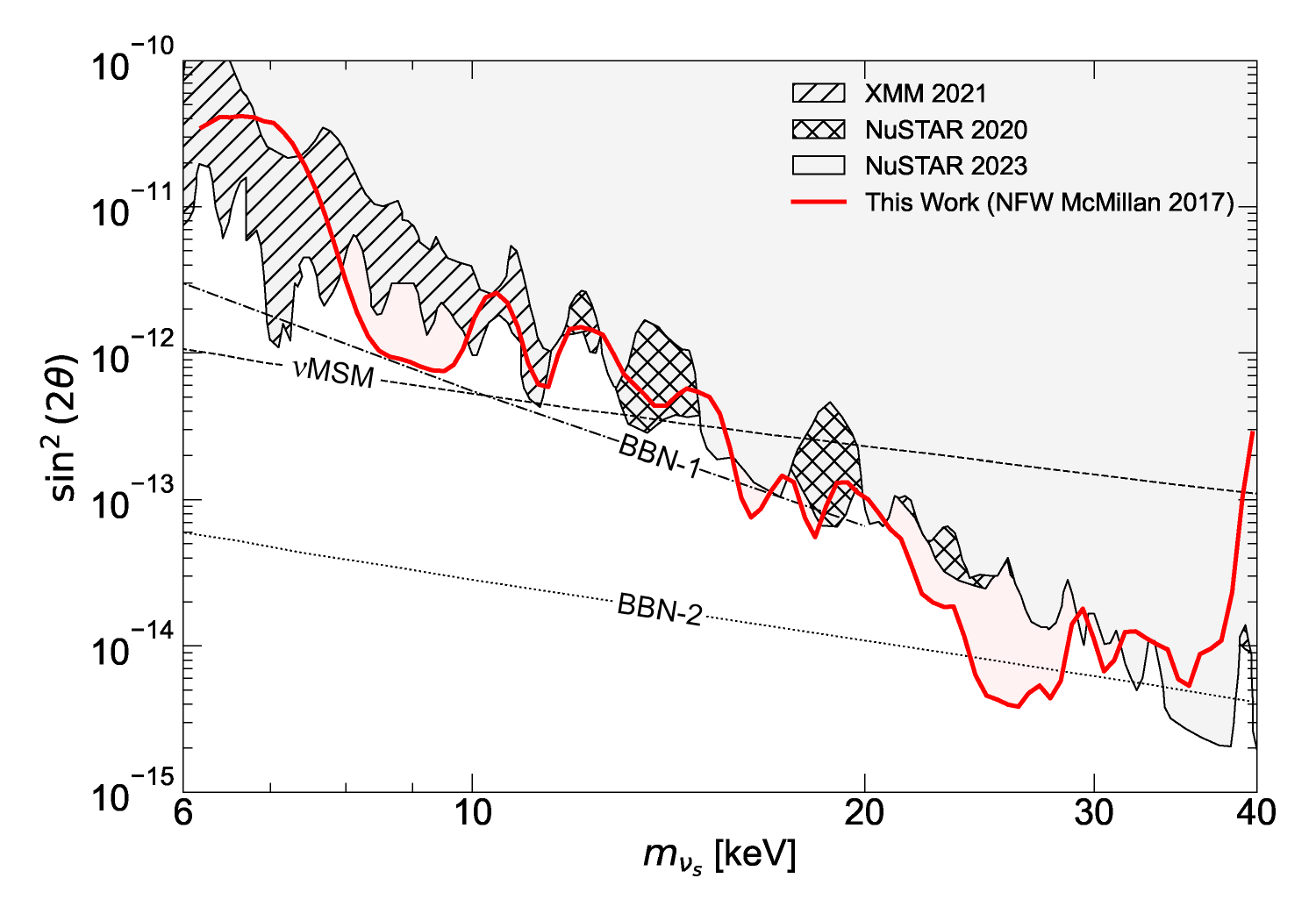}
 %{figures/final_constraints.pdf}
	\caption{95\% upper limits on $sin^2(2\theta)$ derived from the \nustar\ unfocused stray light spectrum for reference profile~\cite{McMillan2017} along with some previous constraints \cite[\texttt{NuSTAR 2020, 2023}:][]{Roach:2022lgo}, \cite[\texttt{XMM 2021}:][]{XMM2021} and updated theoretical lower bounds\,\cite{Canetti:2012kh,Laine:2008pg,Cherry:2017dwu}. Note that the borderline excess at 40\,keV is the known instrumental effect caused by absorbed stray light \cite{2017JATIS...3d4003M,Rossland:2023vfc,Weng:2024rnt}.}
	\label{fig:final_c}
\end{figure}

\begin{figure*}[!htb]
  \centering
  \subfigure{\includegraphics[scale=0.35]{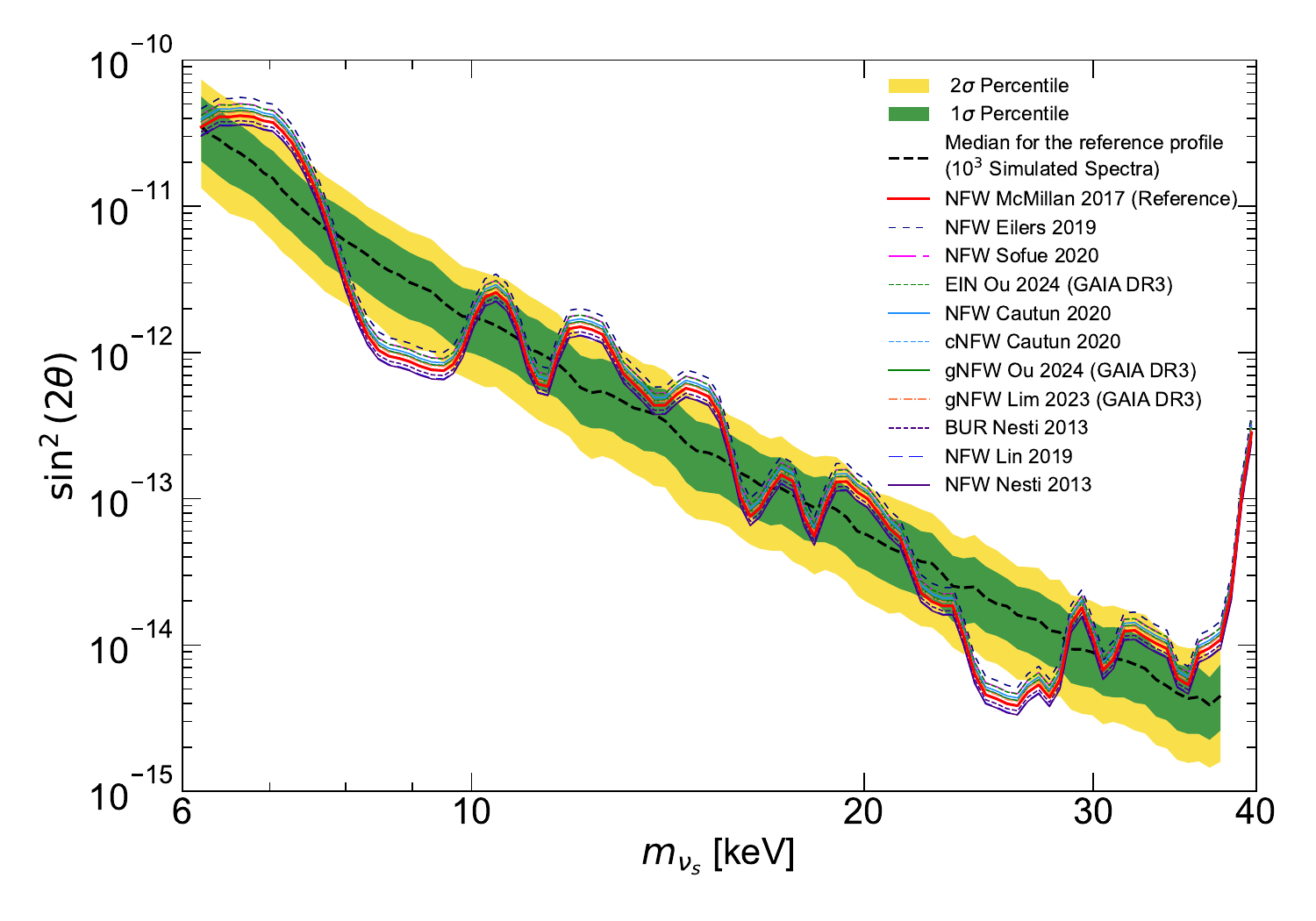}}
  % \hspace*{-1.0cm}
  \subfigure{\includegraphics[scale=0.35]{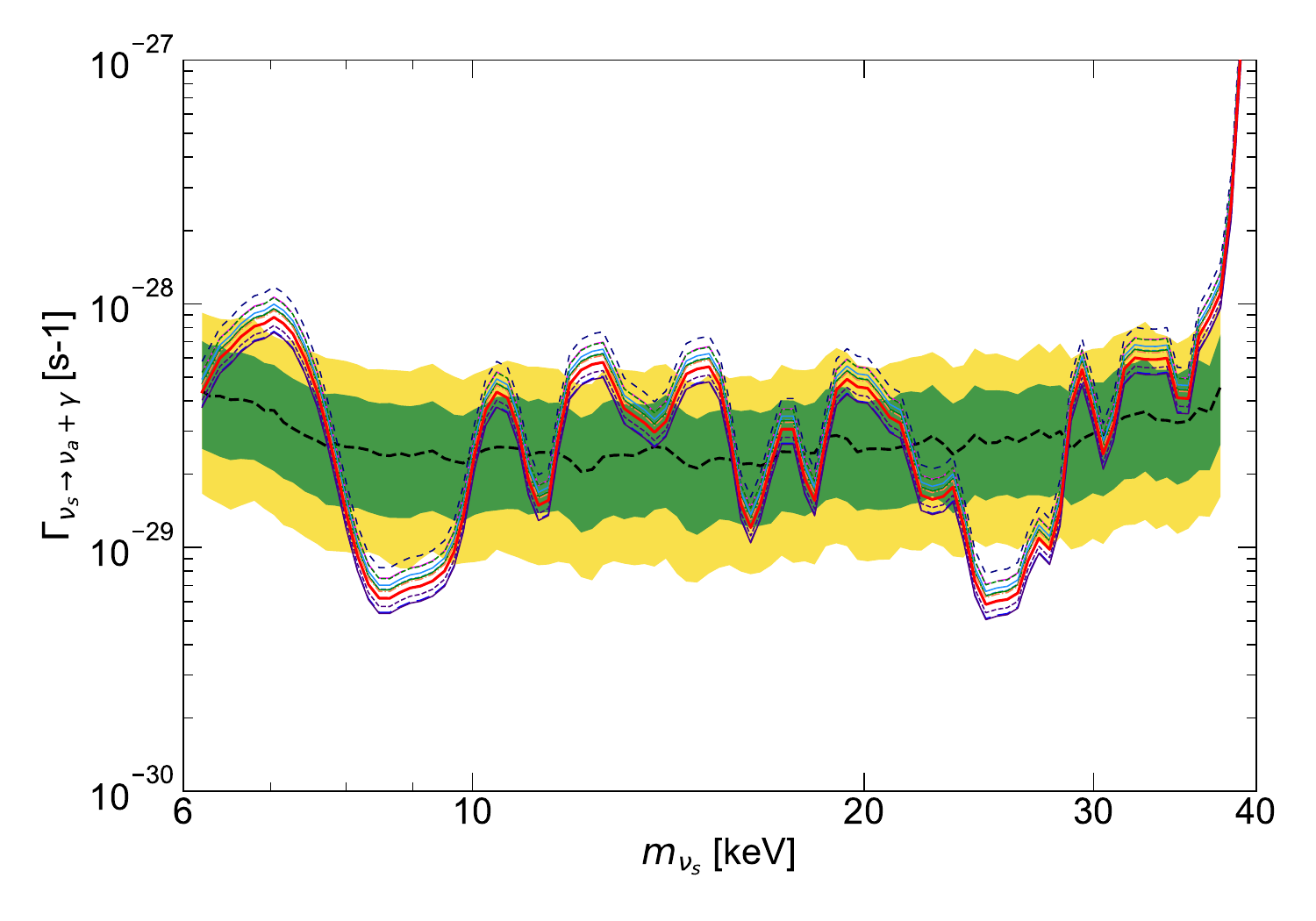}}
  \caption{Upper limits (95\% CL) on the parameters of decaying dark matter. \textbf{Left:} bounds on $sin^2(2\theta)$ derived from the \acxb\ stray light spectrum for different dark matter profiles. The black dashed line shows the median value for the {\it expected limit} obtained with analysis of $10^3$ artificial \acxb-like spectra for DM profile~\cite{McMillan2017}. The green and yellow areas correspond to the 1$\sigma$ and 2$\sigma$ percentiles. \textbf{Right:} the same, but for the width of the decay of sterile neutrinos into active neutrinos and photons. It can be straightforwardly applied to other decaying dark matter models.}
  \label{fig:constrs}
\end{figure*}

We perform a search for this signature in the data of the \nustar\ telescope obtained over 11 years of operation. Our analysis is fully based on the method developed by \cite{Krivonos2021}, where the data from the \nustar\ extragalactic deep fields were used to measure the X-ray spectrum of the Cosmic X-ray Background (CXB) from 3 to 20 keV. Based on this spectrum, Ref.\,\cite{Roach:2022lgo} disfavored a hint possibly related to ${\sim}7$~keV sterile neutrino decaying into a 3.5~keV photon \cite{Bulbul:2014sua,Boyarsky:2014jta}. Here we use the data collected by \nustar\ all over the sky to search for possible traces of the radiative decay \eqref{decays}, assuming that the sterile neutrinos compose the entire dark matter component of our Galaxy, the Milky Way (MW). Similar to other recent studies \cite{Anderson:2014tza,Neronov:2016wdd,Roach:2022lgo,Zakharov:2023mnp}, we find no solid evidence for a monochromatic line to be associated with decay \eqref{decays} and place new upper limits on the active-sterile mixing angle $\theta$ for sterile neutrino masses $m_s=6\!-\!40$\,keV, which are presented in Fig.\,\ref{fig:final_c}
 % ASKY\_AB\_v4\_poly20\_b3 -- this is technical name, not for publication
along with most recent constraints from previous searches. 
% Our constraints do not depend on the origin of the sterile neutrino, but only on the dark matter profile in the Galaxy, relying on the single Galactic dark matter component (and may be straightforwardly rescaled, if the sterile neutrinos form only a fraction of the dark matter). 
Our constraints do not depend on the origin of the sterile neutrino, but like any constraints on indirect dark matter searches, they depend on: a) the dark matter profile in the Galaxy, relying on the single Galactic dark matter component (and may be straightforwardly rescaled, if the sterile neutrinos form only a fraction of the dark matter) and b) The constraints are subject to statistical fluctuations reflecting the specific realization of the data set (for more details see sec.\,4 and Fig.\,\ref{fig:constrs}). Certain parts of the model parameter space may be favored in particular models with specific mechanisms of the sterile neutrino production operating in the early Universe. {\it To illustrate} the relevance of the obtained limits we show also {\it theoretical lower bounds} inherent in the self-contained $\nu$MSM model\,\cite{Canetti:2012kh} (that simultaneously explains neutrino oscillations, baryon asymmetry and dark matter, see review \cite{Drewes:2013gca}) and {\it generic lower bounds}  induced by the Big Bang Nucleosynthesis upper bounds on the lepton asymmetry in the primordial plasma\,\cite{Escudero:2022okz}, which we updated for two different estimates\,\cite{Laine:2008pg,Cherry:2017dwu}, see details in Supplemental Materials (SuMs).   

%\begin{figure}[!ht]
%    \centering
%    \includegraphics[width=\columnwidth]{constraints_expected.png}
%    \caption{Upper limits on mixing (95\%\,C.L.) from 2 years of data taking at ART-XC in the survey mode are depicted with red solid line. The blue dashed line refers to the expected 2-year sensitivity of ART-XC. Note, that the expected sensitivity does not contain statistical fluctuations observed in real data. Black solid line with gray shading show limits recently obtained with the {\nustar} in the Galactic halo \cite{Roach:2022lgo}. And black dotted line with hatching show limits obtained in Galactic bulge \cite{2020PhRvD.101j3011R} with {\nustar} (see our comment in Sec. 4). The white region is remaining allowed area for sterile neutrinos parameters.  The previous upper limits: Fermi-GBM \cite{FermiGBM}, Suzaku \cite{Suzaku}, XMM 2021 \cite{XMM2021} and Chandra and XMM (old) \cite{CXO_Andromeda, XMM_Draco, CXO_Draco, Watson:2006qb}. And the previous lower limits: Milky Way satellite counts (MW SC) \cite{MWSC} and big bang nucleosynthesis (BBN) limit \cite{BBN1, BBN2}.}
%    \label{fig:results}
%\end{figure}
%\newpage

{\it 2. Observations and data processing.}  
The Nuclear Spectroscopic Telescope Array (\nustar) is a NASA orbital X-ray telescope operating in the range of 3 to 79 keV \cite{NuSTAR:2013yza}. \nustar\ carries out two identical co-aligned telescopes, each containing an independent assembly of X-ray mirrors and a focal-plane detector, referred to as focal plane module (FPM) A and B (FPMA and FPMB). The optical system implements the grazing incidence conical approximation Wolter~I design, where X-ray photons are focused by reflection from upper and lower cones. The field of view (FOV) for these reflected twice photons, determined by the detector size,  is ${\sim}13' {\times} 13'$. 
% Each FPM contains four $32{\times}32$ solid state pixel detector arrays (or ``chips''). The detector chips, containing physical pixels of 0.6~mm in size, determine RAW coordinate system with $64{\times}64$ elements in total. 

The concept of this work is based on analysing the side aperture of the \nustar\ telescope (aka stray light, see SuMs for details), which is the direct illumination of the detector from directions of a few degrees away from the optical axis. The \nustar\ stray light is dominated by the CXB \cite{Wik:2014boa,Krivonos2021,Rossland:2023vfc} for high Galactic latitude observations and by Galactic Ridge X-ray Emission (GRXE) when pointed to the Galactic plane \cite{Revnivtsev:2005rj,Krivonos:2006px,2019ApJ...884..153P}. 

We reduced \nustar\ data following \cite{Krivonos2021} to produce the X-ray spectrum of \nustar\ stray light aperture, which we assume to be dominated by  CXB. This ``aperture'' CXB spectrum is hereafter referred to as \acxb.
%Below we describe details of the data analysis that are relevant for the current work. 
All publicly available \nustar\ observations from July 2012 to January 2024 were initially considered for this work. We removed all observations, those ObsID starts with 2, which means Solar System Objects (e.g. the Sun);  observations of Jupiter performed within DDT (ObsID 90311 and 90313); and observations with exposure time less than 1~ks. As a result, the initial list contains 3917 observations with a total exposure of 166~Ms per FPM.

% Since we are mainly interested in the \nustar\ stray light background, 

Then, we applied a fully automatic wavelet-based method for determining detector area suitable for the \nustar\ stray light analysis, specially developed by \cite{2023JATIS...9d8001M} and implemented as a Python \texttt{nuwavdet}\footnote{\url{http://heagit.cosmos.ru/nustar/nuwavdet}} package. The main idea of \texttt{nuwavdet} is to remove any significant focused X-ray flux from the \nustar\ data, leaving the detector area dominated by the unfocused stray light. We run \texttt{nuwavdet} on all available \nustar\ observations using default parameters. As output, \texttt{nuwavdet} produces the list of detector bad-flagged pixels compatible with the \nustar\ Data Analysis Software. We used this list as input to \texttt{nupipeline} task via \texttt{fpm[a,b]\_userbpfile} parameter to exclude any focused X-ray flux and known artifacts \cite{2017JATIS...3d4003M} from the detector images. In the following, we refer to this clean detector area only. We selected 3248 (FPMA, 150.1~Ms) and 3139 (FPMB, 144.9~Ms) \nustar\ observations with more than 40\% of the usable detector area and statistical quality, expressed in terms of the modified Cash statistics per bin \cite{2017A&A...605A..51K}, $C<1.4$. 

\begin{figure*}[htb]
    \centering
    \includegraphics[width=0.9\textwidth]{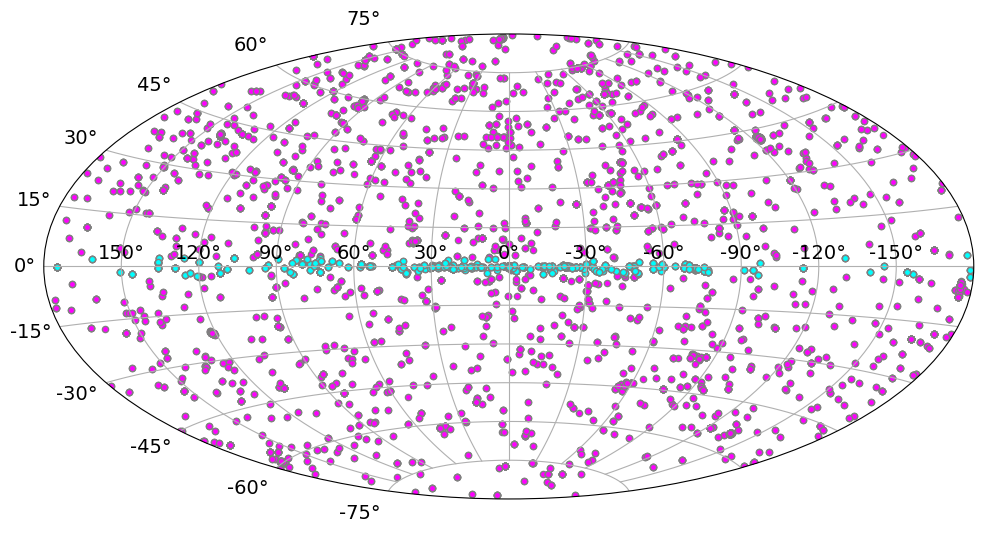}
    \caption{The distribution of 3248 (FPMA) and 3139 (FPMB) \nustar\ observations on the sky in Galactic coordinates. Cyan and magenta points show the \nustar\ observations at $|b|<3^{\circ}$ and $|b|>3^{\circ}$, respectively.}
    \label{fig:obs}
\end{figure*}

The distribution of selected FPMA and FPMB observations over the sky is homogeneous and can be seen in Fig.~\ref{fig:obs} of SuMs, where we also present the long-term evolution of the detector count rate and additional filtering based on it. SuMs also outline the basics of the method and illustrate its validity by applying it to the \nustar\ North Ecliptic Pole (NEP) extragalactic survey. 

% area_max>0.4,  cstat<1.4
% FPMA total exposure 150449.34 ks, total ObsID 3248
% FPMB total exposure 144871.97 ks, total ObsID 3139
% fix_asky_fpmA_gal.csv
% fix_asky_fpmB_gal.csv

% area_max>0.4,  cstat<1.4, bmax>3
% FPMA total exposure 124358.81 ks, total ObsID 2741
% FPMB total exposure 121243.86 ks, total ObsID 2680

% The images were generated in 20 energy intervals logarithmically spaced between 3 and 20 keV. 

\begin{figure}[!t]
    \centering
    \includegraphics[width=\columnwidth]{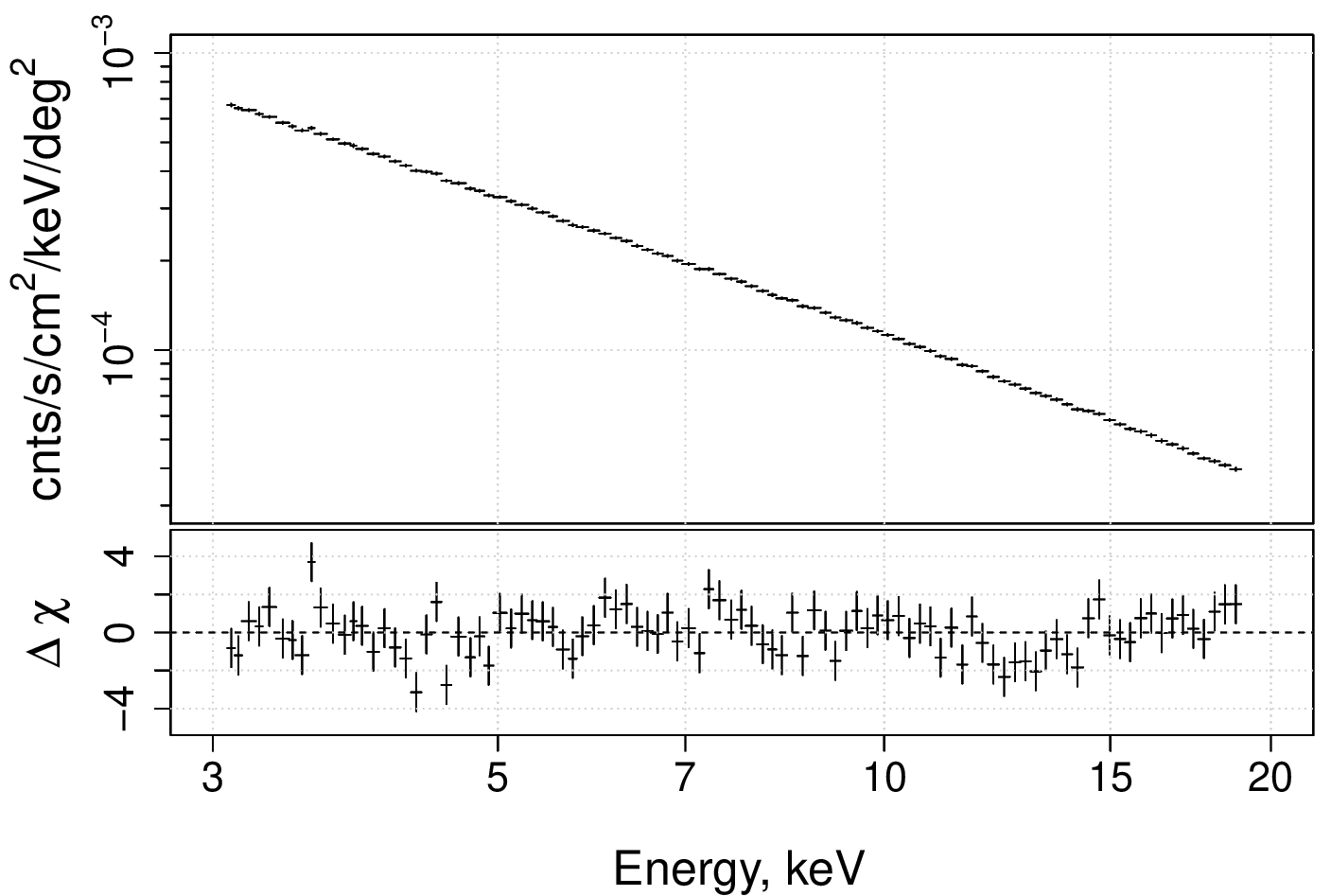}
    \caption{The \nustar\ FPMA and FPMB combined spectrum of the unfocused stray light sky background, measured at $|b|>3^{\circ}$ (\acxb). Due to high statistics in each energy bin, we do not plot the total spectrum model in the upper panel for convenience.}
    \label{fig:egl3}
\end{figure}

\begin{figure*}[tb]
    \centering
    \includegraphics[width=0.9\textwidth]{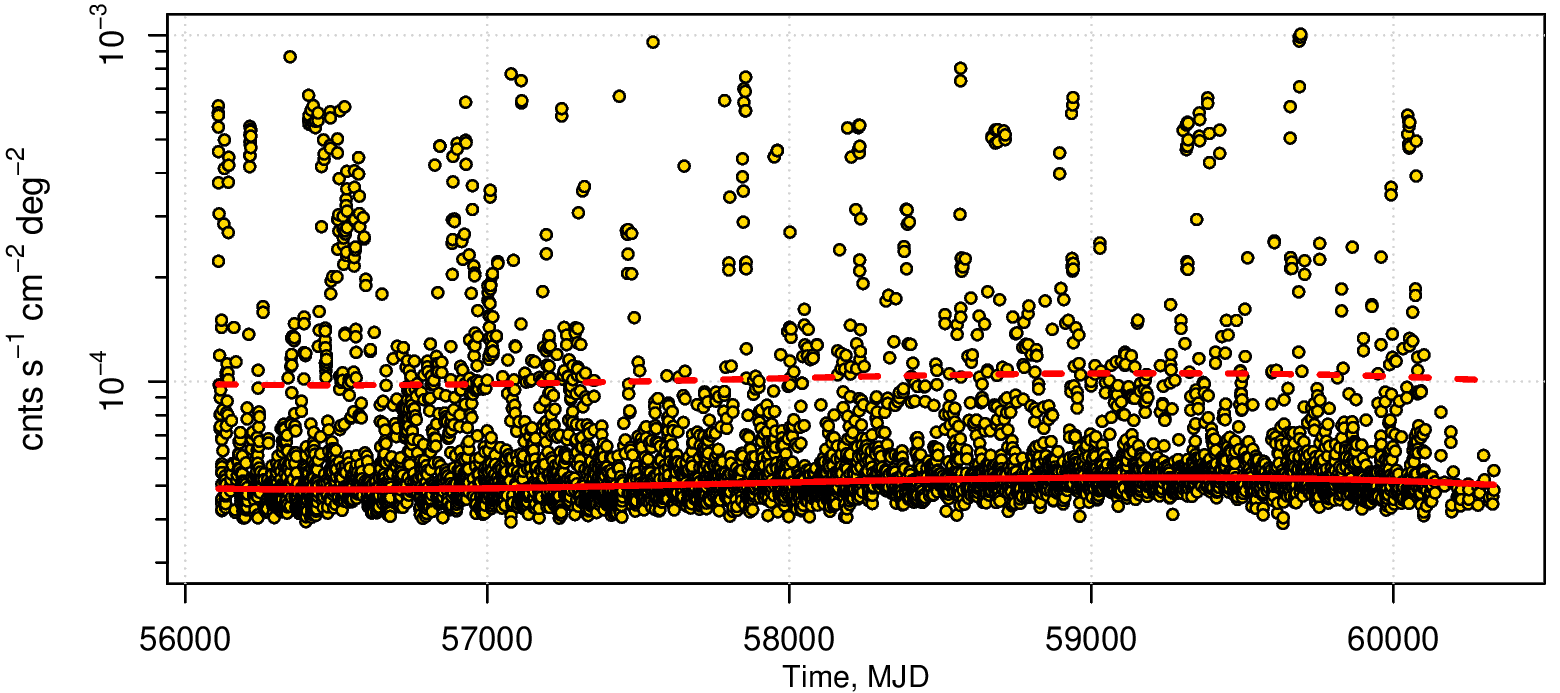}
    \caption{The \nustar\ detector count rate as a function of time. Each point represents individual \nustar\ observation. A solid red line is a cubic polynomial outlier-resistant approximation used to describe long-term variation. A dashed red line shows the same approximation scaled by a factor of 2, that is a threshold: the observations above it are excluded from the analysis.}
    \label{fig:poly}
\end{figure*}

Further we limit our data set to all available observations taken at Galactic latitude $|b|>3^{\circ}$. The total exposure for this data selection, after detector light-curve filtering and combining two FPMs, is 234~Ms (5216 FPMA+FPMB observations). In order to take long-term background variation into account, we modified the original method by adding a relative correction factor of the internal ``flat'' detector component based on cubic polynomial approximation (see Fig.~\ref{fig:poly} of SuMs). The resulting \acxb\ spectrum is shown in Fig.~\ref{fig:egl3}.

Similar to the NEP case described in SuMs, we performed the fitting procedure with fixed $\Gamma_{\rm sol}=4$, $\Gamma_{\rm cxb}$, $E_{\rm cut}$ and allowing normalization parameters and high-energy cutoff to vary. The fit is characterised by $\chi^{2}_{\rm r}$/dof = $130.00/94$ = $1.38$. The high-energy cutoff was estimated at $E_{\rm cut}=34.9\pm0.6$\,keV (hereafter, the uncertainty if fitting parameters is given for 90\% confidence interval). The CXB normalization was measured as $F_{\rm 3-20\,keV}=(3.023\pm0.006)\times 10^{-11}$~\ergscm, which is somewhat higher than measurements \cite{Krivonos2021}, but still consistent, taking overall systematic uncertainty into account. 

Looking at the statistical quality of the \acxb\ spectrum (Fig.~\ref{fig:egl3}), we conclude that the measured data points closely follow a nearly power-law continuum with very low scatter around it as governed by the errors -- a result of a huge exposure. The systematic noise is present, as seen in the residuals and increased $\chi^{2}_{\rm r}$ value, but its contribution is not dominant. We should stress that we did not include any systematics in the spectral fit. For instance, applying 0.5\% systematics improves the fitting quality to $\chi^{2}_{\rm r}$/dof=$89.07/94$ = $0.95$.

{\it 3. Expected signal.} 
The intensity of photons from decays of dark matter particles is written as
\begin{align}
    I_{\gamma}  \equiv \frac{d^2F_{\gamma}}{dE_{\gamma}d\Omega}  \hspace{1ex} \left[\frac{\text{cts}}{\text{cm}^2 \hspace{1pt} \text{s} \hspace{1pt} \text{keV} \hspace{1pt} \text{Sr}}\right],
\end{align}
where $F_{\gamma}$ is the signal flux, $\Omega$ is the solid angle. The signal flux has the following form
\begin{equation}\label{eq:gamma_flux_th_gen}
F_{\gamma} = \frac{1}{4\pi} \frac{\Gamma_{\nu_s}}{m_{\nu_s}} \int \int_{}
 \frac{dN}{dE_{\gamma}} 
 \frac{d\mathcal{D}_{DM}}{d\Omega} dE_{\gamma}d\Omega \hspace{1ex} \left[\frac{\text{cts}}{\text{cm}^2 \hspace{1pt} \text{s} \hspace{1pt}}\right],
\end{equation}
where $\mathcal{D}_{DM}$ is the column density of dark matter along the line of sight in a given solid angle $\Omega$, $dN/dE_{\gamma}$ is the photon spectrum. Since in the two-body decay\,\eqref{decays} the width of the decay line is solely attributed to the Doppler effect bound to MW dark matter particles, it is much smaller than the energy resolution of the telescope, hence, we can set $dN/dE_{\gamma} \sim \delta(E_{\gamma } - m_{\nu_s}/2)$. The $\mathcal{D}_{DM}$ value can be calculated in various ways, taking into account the specific type of dark matter density distribution profile. In our analysis, we consider the spherically symmetric halo, where the dark matter density $\rho_{DM}$ depends only on the radius. Then the quantity $\mathcal{D}_{DM}$ reads
\begin{equation}
\mathcal{D}_{DM} = \int_{f.o.v.} \int_{l.o.s.}
\frac{\rho_{DM}(r)}{z^2} z^2 \text{d}z \text{d} \Omega,
\end{equation}
with integration performed over a given field of view and along the line of sight with the distance to the source $z$. The result of integration over $z$ very mildly depends on the upper limit if it is about the MW virial radius, so we integrate over $z$ in the range $[0,\;2\cdot10^2]$\,kpc. The distance variables are related to the galactic coordinates $(l,b)$ as
\begin{equation}
    r(l, b, z) = \sqrt{z^2 + r_{g.c.}^2 - 2r_{g.c.}z\cos{l}\cos{b}},
\end{equation}
where $r_{g.c.}=8.5$\,kpc is the distance from the observer to the MW center, see Fig~\ref{fig:gal_sph_img} in SuMs for a sketch.

In our analysis, we use stacked spectra for the constraints on the parameters of sterile neutrinos. In this case, the resulting intensity is given by the following expression
\begin{equation}
\label{Intensity}
\begin{aligned}
    \langle I_{\gamma} \rangle & = \frac{1}{4\pi} \frac{\Gamma_{\nu_s}}{m_{\nu_s}} \frac{dN}{dE_{\gamma}} \left[ \frac{1}{\text{T}_{\text{tot}}} \sum_{i} \text{T}_{i} \frac{\text{d} \mathcal{D}_{DM, i}}{\text{d} \Omega} \right]\\
    & \equiv \frac{1}{4\pi} \frac{\Gamma_{\nu_s}}{m_{\nu_s}} \frac{dN}{dE_{\gamma}} \biggl< \frac{\text{d} \mathcal{D}_{DM}}{\text{d} \Omega} \biggr>,
\end{aligned}
\end{equation}
where $\text{T}_{\text{tot}}$ is the total exposure, $\text{T}_{i}$ is the observation time, and $\text{d} \mathcal{D}_{DM, i} / \text{d} \Omega$ is the differential column density of each observation.

To estimate the signal flux, we need to know the distribution of dark matter density in the MW. This quantity is not fixed and is not measured directly, so there are significant uncertainties in estimates of the dark matter distribution, see discussion in SuMs. In the main analysis, we use the dark matter profile from Ref.\,\cite{McMillan2017}.

{\it 4. Spectral Model and Data Analysis.} 
The spectrum fitting is performed with package XSPEC. The \texttt{powerlaw\,+\,cflux(highecut*powerlaw)} model is the base model in our analysis. The first term gives the contribution of solar component of the form
\begin{equation}
    I_{\text{sol}} = N_{\text{sol}}\left(\frac{E_{\gamma}}{1 \text{keV}}\right)^{-\Gamma_{\text{sol}}},
\end{equation}
while the second term gives the contribution of CXB in 3$-$20\,keV band. We use the standard approximation to the CXB component~\cite{Gruber1999}
\begin{equation}
    I_{\text{CXB}} = N_{\rm cxb}\left(\frac{E_{\gamma}}{1 \text{keV}}\right)^{-\Gamma_{\rm cxb}} \text{exp}\left(\frac{E_{\rm cut}^{\rm cxb}-E_{\gamma}}{E_{\text{fold}}}\right),
\end{equation}
and adopt canonical values $N_{\rm cxb} = 2.4 \times 10^{-3}$ and $\Gamma_{\rm cxb} = 1.29$. We set $E_{\rm cut}^{\rm cxb}=10^{-4}$\,keV and choose $E_{\text{fold}}=41.13$\,keV as a trial value. The spectrum intensity and model normalization are expressed in units of cts/cm$^2$/s/keV/deg$^2$. 

We start by fitting {\acxb} spectrum within our base model, that is  without a line from the decay\,\eqref{decays}. Once we found the best fit parameters (see Table.~\ref{tab:bf_pars} of SuMs), we use them as the initial approximation for the entire model, that is the base model augmented by an additive \texttt{gauss} component. To search for constraints on the intensity of the signal from the dark matter decay, we perform a scan placing the center of the gauss component to the center of each energy bin we have in the data sample and carry out the fitting procedure with  free parameters $N_{\text{sol}}$, $\text{lg}_{10}$Flux, $E_{\text{fold}}$ and fixed $\Gamma_{\text{sol}} = 4.0$ as the best fit value (we found no significant changes when it is free).  The minimization procedure yields the intensity of the emission line and the upper limit on the intensity. A one-side upper limit at 95\%CL with one degree of freedom implies $\Delta \chi^2=2.71$.  Thus we bound the intensity\,\eqref{Intensity}, and hence decay rate $\Gamma_{\nu_s}$ and mixing \eqref{eq:neutrino_width}, that we present in Fig.\,\ref{fig:final_c}. 

% \begin{figure}[!htb]
% 	\centering
% 	\includegraphics[width=\columnwidth]{figures/fc_new.pdf}
%  %{figures/final_constraints.pdf}
% 	\caption{95\% upper limits on $sin^2(2\theta)$ derived from the \nustar\ unfocused stray light spectrum for reference profile~\cite{McMillan2017} along with some previous constraints \cite[\texttt{NuSTAR 2020, 2023}:][]{Roach:2022lgo}, \cite[\texttt{XMM 2021}:][]{XMM2021} and updated theoretical lower bounds\,\cite{Canetti:2012kh,Laine:2008pg,Cherry:2017dwu}. Note that the borderline excess at 40\,keV is the known instrumental effect caused by absorbed stray light \citep{2017JATIS...3d4003M,2023AJ....166...20R,2024arXiv240214637W}.}
% 	\label{fig:final_c}
% \end{figure} % ASKY\_AB\_v4\_poly20\_b3 -- this is technical name, not for publication

The procedure of finding upper limits on $sin^2(2\theta)$ is based on a statistical process, which means that the shape of the final constraints will have certain variations since they are a result of a statistical realization, namely, the observed X-ray spectrum. This should be kept in mind when comparing the limits from different studies of this kind. One can distinguish this type of limits as \textit{a given} constraints, in contrast to the \textit{expected} limits, which potentially can be obtained with a given statistics. We assessed our expectations by simulating artificial (based on the best fit parameters) $10^3$ spectra using the \texttt{fakeit} procedure, repeating the line searches and averaging over the obtained limits. The results are presented in Fig.\,\ref{fig:constrs}. One can notice deviations of the observed limits from the expectations. We attribute them to moderate systematics in the observed spectrum, obviously not included in the simulations. This effect can lead to stronger limits in certain mass ranges (e.g. $m_{\nu_s}$=~8$-$9~keV or 23$-$25~keV), where the fluctuations in the observed spectrum go below the spectral model, giving small room for the expected signal.

{\it 5. Conclusion.} The obtained limits are very strong, at the edge of some theoretical predictions:  only a small region remains to be explored  before ruling out the sterile neutrino as a single component dark matter. However, the theoretical predictions obtained with different codes certainly contradict each other. Moreover, there are simple (and physically motivated) modifications of the minimal setup that predict much larger room to explore, see e.g.\,\cite{Bezrukov:2018wvd,Alonso-Alvarez:2021pgy,He:2023neh}. To investigate these models, one definitely needs new instruments, since the presented bounds are based on more than 10 year statistics of high quality, which is difficult to overcome in the nearest future.

\acknowledgements

{\it Acknowledgements.} The work is supported by the RSF Grant No. 22-12-00271. We thank D. Kalashnikov for running the numerical code \texttt{sterile-dm}, J. Beacom, O. Ruchayskiy and M. Shaposhnikov for valuable comments, which helped to improve the article.

%%% REFERENCES
\addcontentsline{toc}{chapter}{\bibname}
\bibliographystyle{apsrev4-1}
\bibliography{refs}

\begin{thebibliography}{52}%
\makeatletter
\providecommand \@ifxundefined [1]{%
 \@ifx{#1\undefined}
}%
\providecommand \@ifnum [1]{%
 \ifnum #1\expandafter \@firstoftwo
 \else \expandafter \@secondoftwo
 \fi
}%
\providecommand \@ifx [1]{%
 \ifx #1\expandafter \@firstoftwo
 \else \expandafter \@secondoftwo
 \fi
}%
\providecommand \natexlab [1]{#1}%
\providecommand \enquote  [1]{``#1''}%
\providecommand \bibnamefont  [1]{#1}%
\providecommand \bibfnamefont [1]{#1}%
\providecommand \citenamefont [1]{#1}%
\providecommand \href@noop [0]{\@secondoftwo}%
\providecommand \href [0]{\begingroup \@sanitize@url \@href}%
\providecommand \@href[1]{\@@startlink{#1}\@@href}%
\providecommand \@@href[1]{\endgroup#1\@@endlink}%
\providecommand \@sanitize@url [0]{\catcode `\\12\catcode `\$12\catcode `\&12\catcode `\#12\catcode `\^12\catcode `\_12\catcode `\%12\relax}%
\providecommand \@@startlink[1]{}%
\providecommand \@@endlink[0]{}%
\providecommand \url  [0]{\begingroup\@sanitize@url \@url }%
\providecommand \@url [1]{\endgroup\@href {#1}{\urlprefix }}%
\providecommand \urlprefix  [0]{URL }%
\providecommand \Eprint [0]{\href }%
\providecommand \doibase [0]{http://dx.doi.org/}%
\providecommand \selectlanguage [0]{\@gobble}%
\providecommand \bibinfo  [0]{\@secondoftwo}%
\providecommand \bibfield  [0]{\@secondoftwo}%
\providecommand \translation [1]{[#1]}%
\providecommand \BibitemOpen [0]{}%
\providecommand \bibitemStop [0]{}%
\providecommand \bibitemNoStop [0]{.\EOS\space}%
\providecommand \EOS [0]{\spacefactor3000\relax}%
\providecommand \BibitemShut  [1]{\csname bibitem#1\endcsname}%
\let\auto@bib@innerbib\@empty
%</preamble>
\bibitem [{\citenamefont {Volkas}(2002)}]{Volkas:2001zb}%
  \BibitemOpen
  \bibfield  {author} {\bibinfo {author} {\bibfnamefont {R.~R.}\ \bibnamefont {Volkas}},\ }\href {\doibase 10.1016/S0146-6410(02)00122-9} {\bibfield  {journal} {\bibinfo  {journal} {Prog. Part. Nucl. Phys.}\ }\textbf {\bibinfo {volume} {48}},\ \bibinfo {pages} {161} (\bibinfo {year} {2002})},\ \Eprint {http://arxiv.org/abs/hep-ph/0111326} {arXiv:hep-ph/0111326} \BibitemShut {NoStop}%
\bibitem [{\citenamefont {Gorbunov}(2014)}]{Gorbunov:2014efa}%
  \BibitemOpen
  \bibfield  {author} {\bibinfo {author} {\bibfnamefont {D.~S.}\ \bibnamefont {Gorbunov}},\ }\href {\doibase 10.3367/UFNe.0184.201405i.0545} {\bibfield  {journal} {\bibinfo  {journal} {Phys. Usp.}\ }\textbf {\bibinfo {volume} {57}},\ \bibinfo {pages} {503} (\bibinfo {year} {2014})}\BibitemShut {NoStop}%
\bibitem [{\citenamefont {Schechter}\ and\ \citenamefont {Valle}(1980)}]{Schechter:1980gr}%
  \BibitemOpen
  \bibfield  {author} {\bibinfo {author} {\bibfnamefont {J.}~\bibnamefont {Schechter}}\ and\ \bibinfo {author} {\bibfnamefont {J.~W.~F.}\ \bibnamefont {Valle}},\ }\href {\doibase 10.1103/PhysRevD.22.2227} {\bibfield  {journal} {\bibinfo  {journal} {Phys. Rev. D}\ }\textbf {\bibinfo {volume} {22}},\ \bibinfo {pages} {2227} (\bibinfo {year} {1980})}\BibitemShut {NoStop}%
\bibitem [{\citenamefont {Shi}\ and\ \citenamefont {Fuller}(1999)}]{Shi:1998km}%
  \BibitemOpen
  \bibfield  {author} {\bibinfo {author} {\bibfnamefont {X.-D.}\ \bibnamefont {Shi}}\ and\ \bibinfo {author} {\bibfnamefont {G.~M.}\ \bibnamefont {Fuller}},\ }\href {\doibase 10.1103/PhysRevLett.82.2832} {\bibfield  {journal} {\bibinfo  {journal} {Phys. Rev. Lett.}\ }\textbf {\bibinfo {volume} {82}},\ \bibinfo {pages} {2832} (\bibinfo {year} {1999})},\ \Eprint {http://arxiv.org/abs/astro-ph/9810076} {arXiv:astro-ph/9810076} \BibitemShut {NoStop}%
\bibitem [{\citenamefont {Pal}\ and\ \citenamefont {Wolfenstein}(1982)}]{PhysRevD.25.766}%
  \BibitemOpen
  \bibfield  {author} {\bibinfo {author} {\bibfnamefont {P.~B.}\ \bibnamefont {Pal}}\ and\ \bibinfo {author} {\bibfnamefont {L.}~\bibnamefont {Wolfenstein}},\ }\href {\doibase 10.1103/PhysRevD.25.766} {\bibfield  {journal} {\bibinfo  {journal} {Phys. Rev. D}\ }\textbf {\bibinfo {volume} {25}},\ \bibinfo {pages} {766} (\bibinfo {year} {1982})}\BibitemShut {NoStop}%
\bibitem [{\citenamefont {Barger}\ \emph {et~al.}(1995)\citenamefont {Barger}, \citenamefont {Phillips},\ and\ \citenamefont {Sarkar}}]{Barger:1995ty}%
  \BibitemOpen
  \bibfield  {author} {\bibinfo {author} {\bibfnamefont {V.~D.}\ \bibnamefont {Barger}}, \bibinfo {author} {\bibfnamefont {R.~J.~N.}\ \bibnamefont {Phillips}}, \ and\ \bibinfo {author} {\bibfnamefont {S.}~\bibnamefont {Sarkar}},\ }\href {\doibase 10.1016/0370-2693(95)00486-5, 10.1016/0370-2693(95)00831-5} {\bibfield  {journal} {\bibinfo  {journal} {Phys. Lett.}\ }\textbf {\bibinfo {volume} {B352}},\ \bibinfo {pages} {365} (\bibinfo {year} {1995})},\ \bibinfo {note} {[Erratum: Phys. Lett.B356,617(1995)]},\ \Eprint {http://arxiv.org/abs/hep-ph/9503295} {arXiv:hep-ph/9503295 [hep-ph]} \BibitemShut {NoStop}%
%%CITATION = HEP-PH/9503295;%%
\bibitem [{\citenamefont {Abazajian}\ \emph {et~al.}(2001)\citenamefont {Abazajian}, \citenamefont {Fuller},\ and\ \citenamefont {Tucker}}]{Abazajian:2001vt}%
  \BibitemOpen
  \bibfield  {author} {\bibinfo {author} {\bibfnamefont {K.}~\bibnamefont {Abazajian}}, \bibinfo {author} {\bibfnamefont {G.~M.}\ \bibnamefont {Fuller}}, \ and\ \bibinfo {author} {\bibfnamefont {W.~H.}\ \bibnamefont {Tucker}},\ }\href {\doibase 10.1086/323867} {\bibfield  {journal} {\bibinfo  {journal} {Astrophys. J.}\ }\textbf {\bibinfo {volume} {562}},\ \bibinfo {pages} {593} (\bibinfo {year} {2001})},\ \Eprint {http://arxiv.org/abs/astro-ph/0106002} {arXiv:astro-ph/0106002 [astro-ph]} \BibitemShut {NoStop}%
%%CITATION = ASTRO-PH/0106002;%%
\bibitem [{\citenamefont {{Krivonos}}\ \emph {et~al.}(2021)\citenamefont {{Krivonos}}, \citenamefont {{Wik}}, \citenamefont {{Grefenstette}}, \citenamefont {{Madsen}}, \citenamefont {{Perez}}, \citenamefont {{Rossland}}, \citenamefont {{Sazonov}},\ and\ \citenamefont {{Zoglauer}}}]{Krivonos2021}%
  \BibitemOpen
  \bibfield  {author} {\bibinfo {author} {\bibfnamefont {R.}~\bibnamefont {{Krivonos}}}, \bibinfo {author} {\bibfnamefont {D.}~\bibnamefont {{Wik}}}, \bibinfo {author} {\bibfnamefont {B.}~\bibnamefont {{Grefenstette}}}, \bibinfo {author} {\bibfnamefont {K.}~\bibnamefont {{Madsen}}}, \bibinfo {author} {\bibfnamefont {K.}~\bibnamefont {{Perez}}}, \bibinfo {author} {\bibfnamefont {S.}~\bibnamefont {{Rossland}}}, \bibinfo {author} {\bibfnamefont {S.}~\bibnamefont {{Sazonov}}}, \ and\ \bibinfo {author} {\bibfnamefont {A.}~\bibnamefont {{Zoglauer}}},\ }\href {\doibase 10.1093/mnras/stab209} {\bibfield  {journal} {\bibinfo  {journal} {Mon. Not. Roy. Astron. Soc.}\ }\textbf {\bibinfo {volume} {502}},\ \bibinfo {pages} {3966} (\bibinfo {year} {2021})},\ \Eprint {http://arxiv.org/abs/2011.11469} {arXiv:2011.11469 [astro-ph.HE]} \BibitemShut {NoStop}%
\bibitem [{\citenamefont {Roach}\ \emph {et~al.}(2023)\citenamefont {Roach}, \citenamefont {Rossland}, \citenamefont {Ng}, \citenamefont {Perez}, \citenamefont {Beacom}, \citenamefont {Grefenstette}, \citenamefont {Horiuchi}, \citenamefont {Krivonos},\ and\ \citenamefont {Wik}}]{Roach:2022lgo}%
  \BibitemOpen
  \bibfield  {author} {\bibinfo {author} {\bibfnamefont {B.~M.}\ \bibnamefont {Roach}}, \bibinfo {author} {\bibfnamefont {S.}~\bibnamefont {Rossland}}, \bibinfo {author} {\bibfnamefont {K.~C.~Y.}\ \bibnamefont {Ng}}, \bibinfo {author} {\bibfnamefont {K.}~\bibnamefont {Perez}}, \bibinfo {author} {\bibfnamefont {J.~F.}\ \bibnamefont {Beacom}}, \bibinfo {author} {\bibfnamefont {B.~W.}\ \bibnamefont {Grefenstette}}, \bibinfo {author} {\bibfnamefont {S.}~\bibnamefont {Horiuchi}}, \bibinfo {author} {\bibfnamefont {R.}~\bibnamefont {Krivonos}}, \ and\ \bibinfo {author} {\bibfnamefont {D.~R.}\ \bibnamefont {Wik}},\ }\href {\doibase 10.1103/PhysRevD.107.023009} {\bibfield  {journal} {\bibinfo  {journal} {Phys. Rev. D}\ }\textbf {\bibinfo {volume} {107}},\ \bibinfo {pages} {023009} (\bibinfo {year} {2023})},\ \Eprint {http://arxiv.org/abs/2207.04572} {arXiv:2207.04572 [astro-ph.HE]} \BibitemShut {NoStop}%
\bibitem [{\citenamefont {Bulbul}\ \emph {et~al.}(2014)\citenamefont {Bulbul}, \citenamefont {Markevitch}, \citenamefont {Foster}, \citenamefont {Smith}, \citenamefont {Loewenstein},\ and\ \citenamefont {Randall}}]{Bulbul:2014sua}%
  \BibitemOpen
  \bibfield  {author} {\bibinfo {author} {\bibfnamefont {E.}~\bibnamefont {Bulbul}}, \bibinfo {author} {\bibfnamefont {M.}~\bibnamefont {Markevitch}}, \bibinfo {author} {\bibfnamefont {A.}~\bibnamefont {Foster}}, \bibinfo {author} {\bibfnamefont {R.~K.}\ \bibnamefont {Smith}}, \bibinfo {author} {\bibfnamefont {M.}~\bibnamefont {Loewenstein}}, \ and\ \bibinfo {author} {\bibfnamefont {S.~W.}\ \bibnamefont {Randall}},\ }\href {\doibase 10.1088/0004-637X/789/1/13} {\bibfield  {journal} {\bibinfo  {journal} {Astrophys. J.}\ }\textbf {\bibinfo {volume} {789}},\ \bibinfo {pages} {13} (\bibinfo {year} {2014})},\ \Eprint {http://arxiv.org/abs/1402.2301} {arXiv:1402.2301 [astro-ph.CO]} \BibitemShut {NoStop}%
\bibitem [{\citenamefont {Boyarsky}\ \emph {et~al.}(2014)\citenamefont {Boyarsky}, \citenamefont {Ruchayskiy}, \citenamefont {Iakubovskyi},\ and\ \citenamefont {Franse}}]{Boyarsky:2014jta}%
  \BibitemOpen
  \bibfield  {author} {\bibinfo {author} {\bibfnamefont {A.}~\bibnamefont {Boyarsky}}, \bibinfo {author} {\bibfnamefont {O.}~\bibnamefont {Ruchayskiy}}, \bibinfo {author} {\bibfnamefont {D.}~\bibnamefont {Iakubovskyi}}, \ and\ \bibinfo {author} {\bibfnamefont {J.}~\bibnamefont {Franse}},\ }\href {\doibase 10.1103/PhysRevLett.113.251301} {\bibfield  {journal} {\bibinfo  {journal} {Phys. Rev. Lett.}\ }\textbf {\bibinfo {volume} {113}},\ \bibinfo {pages} {251301} (\bibinfo {year} {2014})},\ \Eprint {http://arxiv.org/abs/1402.4119} {arXiv:1402.4119 [astro-ph.CO]} \BibitemShut {NoStop}%
\bibitem [{\citenamefont {Anderson}\ \emph {et~al.}(2015)\citenamefont {Anderson}, \citenamefont {Churazov},\ and\ \citenamefont {Bregman}}]{Anderson:2014tza}%
  \BibitemOpen
  \bibfield  {author} {\bibinfo {author} {\bibfnamefont {M.~E.}\ \bibnamefont {Anderson}}, \bibinfo {author} {\bibfnamefont {E.}~\bibnamefont {Churazov}}, \ and\ \bibinfo {author} {\bibfnamefont {J.~N.}\ \bibnamefont {Bregman}},\ }\href {\doibase 10.1093/mnras/stv1559} {\bibfield  {journal} {\bibinfo  {journal} {Mon. Not. Roy. Astron. Soc.}\ }\textbf {\bibinfo {volume} {452}},\ \bibinfo {pages} {3905} (\bibinfo {year} {2015})},\ \Eprint {http://arxiv.org/abs/1408.4115} {arXiv:1408.4115 [astro-ph.HE]} \BibitemShut {NoStop}%
\bibitem [{\citenamefont {Neronov}\ \emph {et~al.}(2016)\citenamefont {Neronov}, \citenamefont {Malyshev},\ and\ \citenamefont {Eckert}}]{Neronov:2016wdd}%
  \BibitemOpen
  \bibfield  {author} {\bibinfo {author} {\bibfnamefont {A.}~\bibnamefont {Neronov}}, \bibinfo {author} {\bibfnamefont {D.}~\bibnamefont {Malyshev}}, \ and\ \bibinfo {author} {\bibfnamefont {D.}~\bibnamefont {Eckert}},\ }\href {\doibase 10.1103/PhysRevD.94.123504} {\bibfield  {journal} {\bibinfo  {journal} {Phys. Rev. D}\ }\textbf {\bibinfo {volume} {94}},\ \bibinfo {pages} {123504} (\bibinfo {year} {2016})},\ \Eprint {http://arxiv.org/abs/1607.07328} {arXiv:1607.07328 [astro-ph.HE]} \BibitemShut {NoStop}%
\bibitem [{\citenamefont {Zakharov}\ \emph {et~al.}(2024)\citenamefont {Zakharov} \emph {et~al.}}]{Zakharov:2023mnp}%
  \BibitemOpen
  \bibfield  {author} {\bibinfo {author} {\bibfnamefont {E.~I.}\ \bibnamefont {Zakharov}} \emph {et~al.},\ }\href {\doibase 10.1103/PhysRevD.109.L021301} {\bibfield  {journal} {\bibinfo  {journal} {Phys. Rev. D}\ }\textbf {\bibinfo {volume} {109}},\ \bibinfo {pages} {L021301} (\bibinfo {year} {2024})},\ \Eprint {http://arxiv.org/abs/2303.12673} {arXiv:2303.12673 [astro-ph.HE]} \BibitemShut {NoStop}%
\bibitem [{\citenamefont {{McMillan}}(2017)}]{McMillan2017}%
  \BibitemOpen
  \bibfield  {author} {\bibinfo {author} {\bibfnamefont {P.~J.}\ \bibnamefont {{McMillan}}},\ }\href {\doibase 10.1093/mnras/stw2759} {\bibfield  {journal} {\bibinfo  {journal} {Mon. Not. Roy. Astron. Soc.}\ }\textbf {\bibinfo {volume} {465}},\ \bibinfo {pages} {76} (\bibinfo {year} {2017})},\ \Eprint {http://arxiv.org/abs/1608.00971} {arXiv:1608.00971 [astro-ph.GA]} \BibitemShut {NoStop}%
\bibitem [{\citenamefont {Foster}\ \emph {et~al.}(2021)\citenamefont {Foster}, \citenamefont {Kongsore}, \citenamefont {Dessert}, \citenamefont {Park}, \citenamefont {Rodd}, \citenamefont {Cranmer},\ and\ \citenamefont {Safdi}}]{XMM2021}%
  \BibitemOpen
  \bibfield  {author} {\bibinfo {author} {\bibfnamefont {J.~W.}\ \bibnamefont {Foster}}, \bibinfo {author} {\bibfnamefont {M.}~\bibnamefont {Kongsore}}, \bibinfo {author} {\bibfnamefont {C.}~\bibnamefont {Dessert}}, \bibinfo {author} {\bibfnamefont {Y.}~\bibnamefont {Park}}, \bibinfo {author} {\bibfnamefont {N.~L.}\ \bibnamefont {Rodd}}, \bibinfo {author} {\bibfnamefont {K.}~\bibnamefont {Cranmer}}, \ and\ \bibinfo {author} {\bibfnamefont {B.~R.}\ \bibnamefont {Safdi}},\ }\href {\doibase 10.1103/PhysRevLett.127.051101} {\bibfield  {journal} {\bibinfo  {journal} {Phys. Rev. Lett.}\ }\textbf {\bibinfo {volume} {127}},\ \bibinfo {pages} {051101} (\bibinfo {year} {2021})},\ \Eprint {http://arxiv.org/abs/2102.02207} {arXiv:2102.02207 [astro-ph.CO]} \BibitemShut {NoStop}%
\bibitem [{\citenamefont {Canetti}\ \emph {et~al.}(2013)\citenamefont {Canetti}, \citenamefont {Drewes}, \citenamefont {Frossard},\ and\ \citenamefont {Shaposhnikov}}]{Canetti:2012kh}%
  \BibitemOpen
  \bibfield  {author} {\bibinfo {author} {\bibfnamefont {L.}~\bibnamefont {Canetti}}, \bibinfo {author} {\bibfnamefont {M.}~\bibnamefont {Drewes}}, \bibinfo {author} {\bibfnamefont {T.}~\bibnamefont {Frossard}}, \ and\ \bibinfo {author} {\bibfnamefont {M.}~\bibnamefont {Shaposhnikov}},\ }\href {\doibase 10.1103/PhysRevD.87.093006} {\bibfield  {journal} {\bibinfo  {journal} {Phys. Rev. D}\ }\textbf {\bibinfo {volume} {87}},\ \bibinfo {pages} {093006} (\bibinfo {year} {2013})},\ \Eprint {http://arxiv.org/abs/1208.4607} {arXiv:1208.4607 [hep-ph]} \BibitemShut {NoStop}%
\bibitem [{\citenamefont {Laine}\ and\ \citenamefont {Shaposhnikov}(2008)}]{Laine:2008pg}%
  \BibitemOpen
  \bibfield  {author} {\bibinfo {author} {\bibfnamefont {M.}~\bibnamefont {Laine}}\ and\ \bibinfo {author} {\bibfnamefont {M.}~\bibnamefont {Shaposhnikov}},\ }\href {\doibase 10.1088/1475-7516/2008/06/031} {\bibfield  {journal} {\bibinfo  {journal} {JCAP}\ }\textbf {\bibinfo {volume} {06}},\ \bibinfo {pages} {031} (\bibinfo {year} {2008})},\ \Eprint {http://arxiv.org/abs/0804.4543} {arXiv:0804.4543 [hep-ph]} \BibitemShut {NoStop}%
\bibitem [{\citenamefont {Cherry}\ and\ \citenamefont {Horiuchi}(2017)}]{Cherry:2017dwu}%
  \BibitemOpen
  \bibfield  {author} {\bibinfo {author} {\bibfnamefont {J.~F.}\ \bibnamefont {Cherry}}\ and\ \bibinfo {author} {\bibfnamefont {S.}~\bibnamefont {Horiuchi}},\ }\href {\doibase 10.1103/PhysRevD.95.083015} {\bibfield  {journal} {\bibinfo  {journal} {Phys. Rev. D}\ }\textbf {\bibinfo {volume} {95}},\ \bibinfo {pages} {083015} (\bibinfo {year} {2017})},\ \Eprint {http://arxiv.org/abs/1701.07874} {arXiv:1701.07874 [hep-ph]} \BibitemShut {NoStop}%
\bibitem [{\citenamefont {{Madsen}}\ \emph {et~al.}(2017)\citenamefont {{Madsen}}, \citenamefont {{Christensen}}, \citenamefont {{Craig}}, \citenamefont {{Forster}}, \citenamefont {{Grefenstette}}, \citenamefont {{Harrison}}, \citenamefont {{Miyasaka}},\ and\ \citenamefont {{Rana}}}]{2017JATIS...3d4003M}%
  \BibitemOpen
  \bibfield  {author} {\bibinfo {author} {\bibfnamefont {K.~K.}\ \bibnamefont {{Madsen}}}, \bibinfo {author} {\bibfnamefont {F.~E.}\ \bibnamefont {{Christensen}}}, \bibinfo {author} {\bibfnamefont {W.~W.}\ \bibnamefont {{Craig}}}, \bibinfo {author} {\bibfnamefont {K.~W.}\ \bibnamefont {{Forster}}}, \bibinfo {author} {\bibfnamefont {B.~W.}\ \bibnamefont {{Grefenstette}}}, \bibinfo {author} {\bibfnamefont {F.~A.}\ \bibnamefont {{Harrison}}}, \bibinfo {author} {\bibfnamefont {H.}~\bibnamefont {{Miyasaka}}}, \ and\ \bibinfo {author} {\bibfnamefont {V.}~\bibnamefont {{Rana}}},\ }\href {\doibase 10.1117/1.JATIS.3.4.044003} {\bibfield  {journal} {\bibinfo  {journal} {Journal of Astronomical Telescopes, Instruments, and Systems}\ }\textbf {\bibinfo {volume} {3}},\ \bibinfo {eid} {044003} (\bibinfo {year} {2017})}\BibitemShut {NoStop}%
\bibitem [{\citenamefont {Rossland}\ \emph {et~al.}(2023)\citenamefont {Rossland} \emph {et~al.}}]{Rossland:2023vfc}%
  \BibitemOpen
  \bibfield  {author} {\bibinfo {author} {\bibfnamefont {S.}~\bibnamefont {Rossland}} \emph {et~al.},\ }\href {\doibase 10.3847/1538-3881/acd0ae} {\bibfield  {journal} {\bibinfo  {journal} {Astron. J.}\ }\textbf {\bibinfo {volume} {166}},\ \bibinfo {pages} {20} (\bibinfo {year} {2023})},\ \Eprint {http://arxiv.org/abs/2304.07962} {arXiv:2304.07962 [astro-ph.HE]} \BibitemShut {NoStop}%
\bibitem [{\citenamefont {Weng}\ \emph {et~al.}(2024)\citenamefont {Weng}, \citenamefont {Zhou}, \citenamefont {Perets}, \citenamefont {Wik},\ and\ \citenamefont {Chen}}]{Weng:2024rnt}%
  \BibitemOpen
  \bibfield  {author} {\bibinfo {author} {\bibfnamefont {J.}~\bibnamefont {Weng}}, \bibinfo {author} {\bibfnamefont {P.}~\bibnamefont {Zhou}}, \bibinfo {author} {\bibfnamefont {H.~B.}\ \bibnamefont {Perets}}, \bibinfo {author} {\bibfnamefont {D.~R.}\ \bibnamefont {Wik}}, \ and\ \bibinfo {author} {\bibfnamefont {Y.}~\bibnamefont {Chen}},\ }\href {\doibase 10.1093/mnras/stae584} {\bibfield  {journal} {\bibinfo  {journal} {Mon. Not. Roy. Astron. Soc.}\ }\textbf {\bibinfo {volume} {529}},\ \bibinfo {pages} {999} (\bibinfo {year} {2024})},\ \Eprint {http://arxiv.org/abs/2402.14637} {arXiv:2402.14637 [astro-ph.HE]} \BibitemShut {NoStop}%
\bibitem [{\citenamefont {Drewes}(2013)}]{Drewes:2013gca}%
  \BibitemOpen
  \bibfield  {author} {\bibinfo {author} {\bibfnamefont {M.}~\bibnamefont {Drewes}},\ }\href {\doibase 10.1142/S0218301313300191} {\bibfield  {journal} {\bibinfo  {journal} {Int. J. Mod. Phys. E}\ }\textbf {\bibinfo {volume} {22}},\ \bibinfo {pages} {1330019} (\bibinfo {year} {2013})},\ \Eprint {http://arxiv.org/abs/1303.6912} {arXiv:1303.6912 [hep-ph]} \BibitemShut {NoStop}%
\bibitem [{\citenamefont {Escudero}\ \emph {et~al.}(2023)\citenamefont {Escudero}, \citenamefont {Ibarra},\ and\ \citenamefont {Maura}}]{Escudero:2022okz}%
  \BibitemOpen
  \bibfield  {author} {\bibinfo {author} {\bibfnamefont {M.}~\bibnamefont {Escudero}}, \bibinfo {author} {\bibfnamefont {A.}~\bibnamefont {Ibarra}}, \ and\ \bibinfo {author} {\bibfnamefont {V.}~\bibnamefont {Maura}},\ }\href {\doibase 10.1103/PhysRevD.107.035024} {\bibfield  {journal} {\bibinfo  {journal} {Phys. Rev. D}\ }\textbf {\bibinfo {volume} {107}},\ \bibinfo {pages} {035024} (\bibinfo {year} {2023})},\ \Eprint {http://arxiv.org/abs/2208.03201} {arXiv:2208.03201 [hep-ph]} \BibitemShut {NoStop}%
\bibitem [{\citenamefont {Harrison}\ \emph {et~al.}(2013)\citenamefont {Harrison} \emph {et~al.}}]{NuSTAR:2013yza}%
  \BibitemOpen
  \bibfield  {author} {\bibinfo {author} {\bibfnamefont {F.~A.}\ \bibnamefont {Harrison}} \emph {et~al.} (\bibinfo {collaboration} {NuSTAR}),\ }\href {\doibase 10.1088/0004-637X/770/2/103} {\bibfield  {journal} {\bibinfo  {journal} {Astrophys. J.}\ }\textbf {\bibinfo {volume} {770}},\ \bibinfo {pages} {103} (\bibinfo {year} {2013})},\ \Eprint {http://arxiv.org/abs/1301.7307} {arXiv:1301.7307 [astro-ph.IM]} \BibitemShut {NoStop}%
\bibitem [{\citenamefont {Wik}\ \emph {et~al.}(2014)\citenamefont {Wik} \emph {et~al.}}]{Wik:2014boa}%
  \BibitemOpen
  \bibfield  {author} {\bibinfo {author} {\bibfnamefont {D.~R.}\ \bibnamefont {Wik}} \emph {et~al.},\ }\href {\doibase 10.1088/0004-637X/792/1/48} {\bibfield  {journal} {\bibinfo  {journal} {Astrophys. J.}\ }\textbf {\bibinfo {volume} {792}},\ \bibinfo {pages} {48} (\bibinfo {year} {2014})},\ \Eprint {http://arxiv.org/abs/1403.2722} {arXiv:1403.2722 [astro-ph.HE]} \BibitemShut {NoStop}%
\bibitem [{\citenamefont {Revnivtsev}\ \emph {et~al.}(2006{\natexlab{a}})\citenamefont {Revnivtsev}, \citenamefont {Sazonov}, \citenamefont {Gilfanov}, \citenamefont {Churazov},\ and\ \citenamefont {Sunyaev}}]{Revnivtsev:2005rj}%
  \BibitemOpen
  \bibfield  {author} {\bibinfo {author} {\bibfnamefont {M.}~\bibnamefont {Revnivtsev}}, \bibinfo {author} {\bibfnamefont {S.}~\bibnamefont {Sazonov}}, \bibinfo {author} {\bibfnamefont {M.}~\bibnamefont {Gilfanov}}, \bibinfo {author} {\bibfnamefont {E.}~\bibnamefont {Churazov}}, \ and\ \bibinfo {author} {\bibfnamefont {R.}~\bibnamefont {Sunyaev}},\ }\href {\doibase 10.1051/0004-6361:20054268} {\bibfield  {journal} {\bibinfo  {journal} {Astron. Astrophys.}\ }\textbf {\bibinfo {volume} {452}},\ \bibinfo {pages} {169} (\bibinfo {year} {2006}{\natexlab{a}})}\BibitemShut {NoStop}%
\bibitem [{\citenamefont {Krivonos}\ \emph {et~al.}(2007)\citenamefont {Krivonos}, \citenamefont {Revnivtsev}, \citenamefont {Churazov}, \citenamefont {Sazonov}, \citenamefont {Grebenev},\ and\ \citenamefont {Sunyaev}}]{Krivonos:2006px}%
  \BibitemOpen
  \bibfield  {author} {\bibinfo {author} {\bibfnamefont {R.}~\bibnamefont {Krivonos}}, \bibinfo {author} {\bibfnamefont {M.}~\bibnamefont {Revnivtsev}}, \bibinfo {author} {\bibfnamefont {E.}~\bibnamefont {Churazov}}, \bibinfo {author} {\bibfnamefont {S.}~\bibnamefont {Sazonov}}, \bibinfo {author} {\bibfnamefont {S.}~\bibnamefont {Grebenev}}, \ and\ \bibinfo {author} {\bibfnamefont {R.}~\bibnamefont {Sunyaev}},\ }\href {\doibase 10.1051/0004-6361:20065626} {\bibfield  {journal} {\bibinfo  {journal} {Astron. Astrophys.}\ }\textbf {\bibinfo {volume} {463}},\ \bibinfo {pages} {957} (\bibinfo {year} {2007})},\ \Eprint {http://arxiv.org/abs/astro-ph/0605420} {arXiv:astro-ph/0605420} \BibitemShut {NoStop}%
\bibitem [{\citenamefont {{Perez}}\ \emph {et~al.}(2019)\citenamefont {{Perez}}, \citenamefont {{Krivonos}},\ and\ \citenamefont {{Wik}}}]{2019ApJ...884..153P}%
  \BibitemOpen
  \bibfield  {author} {\bibinfo {author} {\bibfnamefont {K.}~\bibnamefont {{Perez}}}, \bibinfo {author} {\bibfnamefont {R.}~\bibnamefont {{Krivonos}}}, \ and\ \bibinfo {author} {\bibfnamefont {D.~R.}\ \bibnamefont {{Wik}}},\ }\href {\doibase 10.3847/1538-4357/ab4590} {\bibfield  {journal} {\bibinfo  {journal} {Astrophys. J.}\ }\textbf {\bibinfo {volume} {884}},\ \bibinfo {eid} {153} (\bibinfo {year} {2019})},\ \Eprint {http://arxiv.org/abs/1909.05916} {arXiv:1909.05916 [astro-ph.HE]} \BibitemShut {NoStop}%
\bibitem [{\citenamefont {{Mukhin}}\ \emph {et~al.}(2023)\citenamefont {{Mukhin}}, \citenamefont {{Krivonos}}, \citenamefont {{Vikhlinin}}, \citenamefont {{Grefenstette}}, \citenamefont {{Madsen}},\ and\ \citenamefont {{Wik}}}]{2023JATIS...9d8001M}%
  \BibitemOpen
  \bibfield  {author} {\bibinfo {author} {\bibfnamefont {A.}~\bibnamefont {{Mukhin}}}, \bibinfo {author} {\bibfnamefont {R.}~\bibnamefont {{Krivonos}}}, \bibinfo {author} {\bibfnamefont {A.}~\bibnamefont {{Vikhlinin}}}, \bibinfo {author} {\bibfnamefont {B.}~\bibnamefont {{Grefenstette}}}, \bibinfo {author} {\bibfnamefont {K.}~\bibnamefont {{Madsen}}}, \ and\ \bibinfo {author} {\bibfnamefont {D.}~\bibnamefont {{Wik}}},\ }\href {\doibase 10.1117/1.JATIS.9.4.048001} {\bibfield  {journal} {\bibinfo  {journal} {Journal of Astronomical Telescopes, Instruments, and Systems}\ }\textbf {\bibinfo {volume} {9}},\ \bibinfo {eid} {048001} (\bibinfo {year} {2023})},\ \Eprint {http://arxiv.org/abs/2310.10516} {arXiv:2310.10516 [astro-ph.IM]} \BibitemShut {NoStop}%
\bibitem [{\citenamefont {{Kaastra}}(2017)}]{2017A&A...605A..51K}%
  \BibitemOpen
  \bibfield  {author} {\bibinfo {author} {\bibfnamefont {J.~S.}\ \bibnamefont {{Kaastra}}},\ }\href {\doibase 10.1051/0004-6361/201629319} {\bibfield  {journal} {\bibinfo  {journal} {Astron. Astrophys.}\ }\textbf {\bibinfo {volume} {605}},\ \bibinfo {eid} {A51} (\bibinfo {year} {2017})},\ \Eprint {http://arxiv.org/abs/1707.09202} {arXiv:1707.09202 [astro-ph.HE]} \BibitemShut {NoStop}%
\bibitem [{\citenamefont {{Gruber}}\ \emph {et~al.}(1999{\natexlab{a}})\citenamefont {{Gruber}}, \citenamefont {{Matteson}}, \citenamefont {{Peterson}},\ and\ \citenamefont {{Jung}}}]{Gruber1999}%
  \BibitemOpen
  \bibfield  {author} {\bibinfo {author} {\bibfnamefont {D.~E.}\ \bibnamefont {{Gruber}}}, \bibinfo {author} {\bibfnamefont {J.~L.}\ \bibnamefont {{Matteson}}}, \bibinfo {author} {\bibfnamefont {L.~E.}\ \bibnamefont {{Peterson}}}, \ and\ \bibinfo {author} {\bibfnamefont {G.~V.}\ \bibnamefont {{Jung}}},\ }\href {\doibase 10.1086/307450} {\bibfield  {journal} {\bibinfo  {journal} {Astrophys. J.}\ }\textbf {\bibinfo {volume} {520}},\ \bibinfo {pages} {124} (\bibinfo {year} {1999}{\natexlab{a}})},\ \Eprint {http://arxiv.org/abs/astro-ph/9903492} {arXiv:astro-ph/9903492 [astro-ph]} \BibitemShut {NoStop}%
\bibitem [{\citenamefont {Bezrukov}\ \emph {et~al.}(2019)\citenamefont {Bezrukov}, \citenamefont {Chudaykin},\ and\ \citenamefont {Gorbunov}}]{Bezrukov:2018wvd}%
  \BibitemOpen
  \bibfield  {author} {\bibinfo {author} {\bibfnamefont {F.}~\bibnamefont {Bezrukov}}, \bibinfo {author} {\bibfnamefont {A.}~\bibnamefont {Chudaykin}}, \ and\ \bibinfo {author} {\bibfnamefont {D.}~\bibnamefont {Gorbunov}},\ }\href {\doibase 10.1103/PhysRevD.99.083507} {\bibfield  {journal} {\bibinfo  {journal} {Phys. Rev. D}\ }\textbf {\bibinfo {volume} {99}},\ \bibinfo {pages} {083507} (\bibinfo {year} {2019})},\ \Eprint {http://arxiv.org/abs/1809.09123} {arXiv:1809.09123 [hep-ph]} \BibitemShut {NoStop}%
\bibitem [{\citenamefont {Alonso-\'Alvarez}\ and\ \citenamefont {Cline}(2021)}]{Alonso-Alvarez:2021pgy}%
  \BibitemOpen
  \bibfield  {author} {\bibinfo {author} {\bibfnamefont {G.}~\bibnamefont {Alonso-\'Alvarez}}\ and\ \bibinfo {author} {\bibfnamefont {J.~M.}\ \bibnamefont {Cline}},\ }\href {\doibase 10.1088/1475-7516/2021/10/041} {\bibfield  {journal} {\bibinfo  {journal} {JCAP}\ }\textbf {\bibinfo {volume} {10}},\ \bibinfo {pages} {041} (\bibinfo {year} {2021})},\ \Eprint {http://arxiv.org/abs/2107.07524} {arXiv:2107.07524 [hep-ph]} \BibitemShut {NoStop}%
\bibitem [{\citenamefont {He}\ \emph {et~al.}(2023)\citenamefont {He}, \citenamefont {Liu}, \citenamefont {Ma},\ and\ \citenamefont {Wang}}]{He:2023neh}%
  \BibitemOpen
  \bibfield  {author} {\bibinfo {author} {\bibfnamefont {Y.}~\bibnamefont {He}}, \bibinfo {author} {\bibfnamefont {J.}~\bibnamefont {Liu}}, \bibinfo {author} {\bibfnamefont {X.}~\bibnamefont {Ma}}, \ and\ \bibinfo {author} {\bibfnamefont {X.-P.}\ \bibnamefont {Wang}},\ }\href {\doibase 10.1088/1475-7516/2023/09/047} {\bibfield  {journal} {\bibinfo  {journal} {JCAP}\ }\textbf {\bibinfo {volume} {09}},\ \bibinfo {pages} {047} (\bibinfo {year} {2023})},\ \Eprint {http://arxiv.org/abs/2305.08095} {arXiv:2305.08095 [hep-ph]} \BibitemShut {NoStop}%
\bibitem [{\citenamefont {Venumadhav}\ \emph {et~al.}(2016)\citenamefont {Venumadhav}, \citenamefont {Cyr-Racine}, \citenamefont {Abazajian},\ and\ \citenamefont {Hirata}}]{Venumadhav:2015pla}%
  \BibitemOpen
  \bibfield  {author} {\bibinfo {author} {\bibfnamefont {T.}~\bibnamefont {Venumadhav}}, \bibinfo {author} {\bibfnamefont {F.-Y.}\ \bibnamefont {Cyr-Racine}}, \bibinfo {author} {\bibfnamefont {K.~N.}\ \bibnamefont {Abazajian}}, \ and\ \bibinfo {author} {\bibfnamefont {C.~M.}\ \bibnamefont {Hirata}},\ }\href {\doibase 10.1103/PhysRevD.94.043515} {\bibfield  {journal} {\bibinfo  {journal} {Phys. Rev. D}\ }\textbf {\bibinfo {volume} {94}},\ \bibinfo {pages} {043515} (\bibinfo {year} {2016})},\ \Eprint {http://arxiv.org/abs/1507.06655} {arXiv:1507.06655 [astro-ph.CO]} \BibitemShut {NoStop}%
\bibitem [{\citenamefont {Ghiglieri}\ and\ \citenamefont {Laine}(2015)}]{Ghiglieri:2015jua}%
  \BibitemOpen
  \bibfield  {author} {\bibinfo {author} {\bibfnamefont {J.}~\bibnamefont {Ghiglieri}}\ and\ \bibinfo {author} {\bibfnamefont {M.}~\bibnamefont {Laine}},\ }\href {\doibase 10.1007/JHEP11(2015)171} {\bibfield  {journal} {\bibinfo  {journal} {JHEP}\ }\textbf {\bibinfo {volume} {11}},\ \bibinfo {pages} {171} (\bibinfo {year} {2015})},\ \Eprint {http://arxiv.org/abs/1506.06752} {arXiv:1506.06752 [hep-ph]} \BibitemShut {NoStop}%
\bibitem [{\citenamefont {Asaka}\ \emph {et~al.}(2005)\citenamefont {Asaka}, \citenamefont {Blanchet},\ and\ \citenamefont {Shaposhnikov}}]{Asaka:2005an}%
  \BibitemOpen
  \bibfield  {author} {\bibinfo {author} {\bibfnamefont {T.}~\bibnamefont {Asaka}}, \bibinfo {author} {\bibfnamefont {S.}~\bibnamefont {Blanchet}}, \ and\ \bibinfo {author} {\bibfnamefont {M.}~\bibnamefont {Shaposhnikov}},\ }\href {\doibase 10.1016/j.physletb.2005.09.070} {\bibfield  {journal} {\bibinfo  {journal} {Phys. Lett. B}\ }\textbf {\bibinfo {volume} {631}},\ \bibinfo {pages} {151} (\bibinfo {year} {2005})},\ \Eprint {http://arxiv.org/abs/hep-ph/0503065} {arXiv:hep-ph/0503065} \BibitemShut {NoStop}%
\bibitem [{\citenamefont {Asaka}\ and\ \citenamefont {Shaposhnikov}(2005)}]{Asaka:2005pn}%
  \BibitemOpen
  \bibfield  {author} {\bibinfo {author} {\bibfnamefont {T.}~\bibnamefont {Asaka}}\ and\ \bibinfo {author} {\bibfnamefont {M.}~\bibnamefont {Shaposhnikov}},\ }\href {\doibase 10.1016/j.physletb.2005.06.020} {\bibfield  {journal} {\bibinfo  {journal} {Phys. Lett. B}\ }\textbf {\bibinfo {volume} {620}},\ \bibinfo {pages} {17} (\bibinfo {year} {2005})},\ \Eprint {http://arxiv.org/abs/hep-ph/0505013} {arXiv:hep-ph/0505013} \BibitemShut {NoStop}%
\bibitem [{\citenamefont {Boyarsky}\ \emph {et~al.}(2009)\citenamefont {Boyarsky}, \citenamefont {Ruchayskiy},\ and\ \citenamefont {Shaposhnikov}}]{Boyarsky:2009ix}%
  \BibitemOpen
  \bibfield  {author} {\bibinfo {author} {\bibfnamefont {A.}~\bibnamefont {Boyarsky}}, \bibinfo {author} {\bibfnamefont {O.}~\bibnamefont {Ruchayskiy}}, \ and\ \bibinfo {author} {\bibfnamefont {M.}~\bibnamefont {Shaposhnikov}},\ }\href {\doibase 10.1146/annurev.nucl.010909.083654} {\bibfield  {journal} {\bibinfo  {journal} {Ann. Rev. Nucl. Part. Sci.}\ }\textbf {\bibinfo {volume} {59}},\ \bibinfo {pages} {191} (\bibinfo {year} {2009})},\ \Eprint {http://arxiv.org/abs/0901.0011} {arXiv:0901.0011 [hep-ph]} \BibitemShut {NoStop}%
\bibitem [{\citenamefont {Perez}\ \emph {et~al.}(2019)\citenamefont {Perez}, \citenamefont {Krivonos},\ and\ \citenamefont {Wik}}]{Perez:2019bgo}%
  \BibitemOpen
  \bibfield  {author} {\bibinfo {author} {\bibfnamefont {K.}~\bibnamefont {Perez}}, \bibinfo {author} {\bibfnamefont {R.}~\bibnamefont {Krivonos}}, \ and\ \bibinfo {author} {\bibfnamefont {D.~R.}\ \bibnamefont {Wik}},\ }\href {\doibase 10.3847/1538-4357/ab4590} {\  (\bibinfo {year} {2019}),\ 10.3847/1538-4357/ab4590},\ \Eprint {http://arxiv.org/abs/1909.05916} {arXiv:1909.05916 [astro-ph.HE]} \BibitemShut {NoStop}%
\bibitem [{\citenamefont {Zhao}\ \emph {et~al.}(2024)\citenamefont {Zhao} \emph {et~al.}}]{Zhao:2024klp}%
  \BibitemOpen
  \bibfield  {author} {\bibinfo {author} {\bibfnamefont {X.}~\bibnamefont {Zhao}} \emph {et~al.},\ }\href {\doibase 10.3847/1538-4357/ad2b61} {\bibfield  {journal} {\bibinfo  {journal} {Astrophys. J.}\ }\textbf {\bibinfo {volume} {965}},\ \bibinfo {pages} {188} (\bibinfo {year} {2024})},\ \Eprint {http://arxiv.org/abs/2402.13508} {arXiv:2402.13508 [astro-ph.HE]} \BibitemShut {NoStop}%
\bibitem [{\citenamefont {Windhorst}\ \emph {et~al.}(2023)\citenamefont {Windhorst} \emph {et~al.}}]{Windhorst:2022rrv}%
  \BibitemOpen
  \bibfield  {author} {\bibinfo {author} {\bibfnamefont {R.~A.}\ \bibnamefont {Windhorst}} \emph {et~al.},\ }\href {\doibase 10.3847/1538-3881/aca163} {\bibfield  {journal} {\bibinfo  {journal} {Astron. J.}\ }\textbf {\bibinfo {volume} {165}},\ \bibinfo {pages} {13} (\bibinfo {year} {2023})},\ \Eprint {http://arxiv.org/abs/2209.04119} {arXiv:2209.04119 [astro-ph.CO]} \BibitemShut {NoStop}%
\bibitem [{\citenamefont {{Gruber}}\ \emph {et~al.}(1999{\natexlab{b}})\citenamefont {{Gruber}}, \citenamefont {{Matteson}}, \citenamefont {{Peterson}},\ and\ \citenamefont {{Jung}}}]{1999ApJ...520..124G}%
  \BibitemOpen
  \bibfield  {author} {\bibinfo {author} {\bibfnamefont {D.~E.}\ \bibnamefont {{Gruber}}}, \bibinfo {author} {\bibfnamefont {J.~L.}\ \bibnamefont {{Matteson}}}, \bibinfo {author} {\bibfnamefont {L.~E.}\ \bibnamefont {{Peterson}}}, \ and\ \bibinfo {author} {\bibfnamefont {G.~V.}\ \bibnamefont {{Jung}}},\ }\href {\doibase 10.1086/307450} {\bibfield  {journal} {\bibinfo  {journal} {Astrophys. J.}\ }\textbf {\bibinfo {volume} {520}},\ \bibinfo {pages} {124} (\bibinfo {year} {1999}{\natexlab{b}})},\ \Eprint {http://arxiv.org/abs/astro-ph/9903492} {arXiv:astro-ph/9903492 [astro-ph]} \BibitemShut {NoStop}%
\bibitem [{\citenamefont {Revnivtsev}\ \emph {et~al.}(2006{\natexlab{b}})\citenamefont {Revnivtsev}, \citenamefont {Molkov},\ and\ \citenamefont {Sazonov}}]{Revnivtsev:2006gf}%
  \BibitemOpen
  \bibfield  {author} {\bibinfo {author} {\bibfnamefont {M.}~\bibnamefont {Revnivtsev}}, \bibinfo {author} {\bibfnamefont {S.}~\bibnamefont {Molkov}}, \ and\ \bibinfo {author} {\bibfnamefont {S.}~\bibnamefont {Sazonov}},\ }\href {\doibase 10.1111/j.1745-3933.2006.00233.x} {\bibfield  {journal} {\bibinfo  {journal} {Mon. Not. Roy. Astron. Soc.}\ }\textbf {\bibinfo {volume} {373}},\ \bibinfo {pages} {L11} (\bibinfo {year} {2006}{\natexlab{b}})},\ \Eprint {http://arxiv.org/abs/astro-ph/0605693} {arXiv:astro-ph/0605693} \BibitemShut {NoStop}%
\bibitem [{\citenamefont {Cautun}\ \emph {et~al.}(2020)\citenamefont {Cautun}, \citenamefont {Benitez-Llambay}, \citenamefont {Deason}, \citenamefont {Frenk}, \citenamefont {Fattahi}, \citenamefont {G\'omez}, \citenamefont {Grand}, \citenamefont {Oman}, \citenamefont {Navarro},\ and\ \citenamefont {Simpson}}]{Cautun:2019eaf}%
  \BibitemOpen
  \bibfield  {author} {\bibinfo {author} {\bibfnamefont {M.}~\bibnamefont {Cautun}}, \bibinfo {author} {\bibfnamefont {A.}~\bibnamefont {Benitez-Llambay}}, \bibinfo {author} {\bibfnamefont {A.~J.}\ \bibnamefont {Deason}}, \bibinfo {author} {\bibfnamefont {C.~S.}\ \bibnamefont {Frenk}}, \bibinfo {author} {\bibfnamefont {A.}~\bibnamefont {Fattahi}}, \bibinfo {author} {\bibfnamefont {F.~A.}\ \bibnamefont {G\'omez}}, \bibinfo {author} {\bibfnamefont {R.~J.~J.}\ \bibnamefont {Grand}}, \bibinfo {author} {\bibfnamefont {K.~A.}\ \bibnamefont {Oman}}, \bibinfo {author} {\bibfnamefont {J.~F.}\ \bibnamefont {Navarro}}, \ and\ \bibinfo {author} {\bibfnamefont {C.~M.}\ \bibnamefont {Simpson}},\ }\href {\doibase 10.1093/mnras/staa1017} {\bibfield  {journal} {\bibinfo  {journal} {Mon. Not. Roy. Astron. Soc.}\ }\textbf {\bibinfo {volume} {494}},\ \bibinfo {pages} {4291} (\bibinfo {year} {2020})},\ \Eprint {http://arxiv.org/abs/1911.04557} {arXiv:1911.04557 [astro-ph.GA]} \BibitemShut {NoStop}%
\bibitem [{\citenamefont {{Nesti}}\ and\ \citenamefont {{Salucci}}(2013)}]{Nesti2013}%
  \BibitemOpen
  \bibfield  {author} {\bibinfo {author} {\bibfnamefont {F.}~\bibnamefont {{Nesti}}}\ and\ \bibinfo {author} {\bibfnamefont {P.}~\bibnamefont {{Salucci}}},\ }\href {\doibase 10.1088/1475-7516/2013/07/016} {\bibfield  {journal} {\bibinfo  {journal} {\jcap}\ }\textbf {\bibinfo {volume} {2013}},\ \bibinfo {eid} {016} (\bibinfo {year} {2013})},\ \Eprint {http://arxiv.org/abs/1304.5127} {arXiv:1304.5127 [astro-ph.GA]} \BibitemShut {NoStop}%
\bibitem [{\citenamefont {{Ou}}\ \emph {et~al.}(2024)\citenamefont {{Ou}}, \citenamefont {{Eilers}}, \citenamefont {{Necib}},\ and\ \citenamefont {{Frebel}}}]{Ou2024}%
  \BibitemOpen
  \bibfield  {author} {\bibinfo {author} {\bibfnamefont {X.}~\bibnamefont {{Ou}}}, \bibinfo {author} {\bibfnamefont {A.-C.}\ \bibnamefont {{Eilers}}}, \bibinfo {author} {\bibfnamefont {L.}~\bibnamefont {{Necib}}}, \ and\ \bibinfo {author} {\bibfnamefont {A.}~\bibnamefont {{Frebel}}},\ }\href {\doibase 10.1093/mnras/stae034} {\bibfield  {journal} {\bibinfo  {journal} {Mon. Not. Roy. Astron. Soc.}\ }\textbf {\bibinfo {volume} {528}},\ \bibinfo {pages} {693} (\bibinfo {year} {2024})},\ \Eprint {http://arxiv.org/abs/2303.12838} {arXiv:2303.12838 [astro-ph.GA]} \BibitemShut {NoStop}%
\bibitem [{\citenamefont {{Lim}}\ \emph {et~al.}(2023)\citenamefont {{Lim}}, \citenamefont {{Putney}}, \citenamefont {{Buckley}},\ and\ \citenamefont {{Shih}}}]{Lim2023}%
  \BibitemOpen
  \bibfield  {author} {\bibinfo {author} {\bibfnamefont {S.~H.}\ \bibnamefont {{Lim}}}, \bibinfo {author} {\bibfnamefont {E.}~\bibnamefont {{Putney}}}, \bibinfo {author} {\bibfnamefont {M.~R.}\ \bibnamefont {{Buckley}}}, \ and\ \bibinfo {author} {\bibfnamefont {D.}~\bibnamefont {{Shih}}},\ }\href {\doibase 10.48550/arXiv.2305.13358} {\bibfield  {journal} {\bibinfo  {journal} {arXiv e-prints}\ ,\ \bibinfo {eid} {arXiv:2305.13358}} (\bibinfo {year} {2023})},\ \Eprint {http://arxiv.org/abs/2305.13358} {arXiv:2305.13358 [astro-ph.GA]} \BibitemShut {NoStop}%
\bibitem [{\citenamefont {{Eilers}}\ \emph {et~al.}(2019)\citenamefont {{Eilers}}, \citenamefont {{Hogg}}, \citenamefont {{Rix}},\ and\ \citenamefont {{Ness}}}]{Eilers2019}%
  \BibitemOpen
  \bibfield  {author} {\bibinfo {author} {\bibfnamefont {A.-C.}\ \bibnamefont {{Eilers}}}, \bibinfo {author} {\bibfnamefont {D.~W.}\ \bibnamefont {{Hogg}}}, \bibinfo {author} {\bibfnamefont {H.-W.}\ \bibnamefont {{Rix}}}, \ and\ \bibinfo {author} {\bibfnamefont {M.~K.}\ \bibnamefont {{Ness}}},\ }\href {\doibase 10.3847/1538-4357/aaf648} {\bibfield  {journal} {\bibinfo  {journal} {Astrophys. J.}\ }\textbf {\bibinfo {volume} {871}},\ \bibinfo {eid} {120} (\bibinfo {year} {2019})},\ \Eprint {http://arxiv.org/abs/1810.09466} {arXiv:1810.09466 [astro-ph.GA]} \BibitemShut {NoStop}%
\bibitem [{\citenamefont {{Lin}}\ and\ \citenamefont {{Li}}(2019)}]{Lin2019}%
  \BibitemOpen
  \bibfield  {author} {\bibinfo {author} {\bibfnamefont {H.-N.}\ \bibnamefont {{Lin}}}\ and\ \bibinfo {author} {\bibfnamefont {X.}~\bibnamefont {{Li}}},\ }\href {\doibase 10.1093/mnras/stz1698} {\bibfield  {journal} {\bibinfo  {journal} {Mon. Not. Roy. Astron. Soc.}\ }\textbf {\bibinfo {volume} {487}},\ \bibinfo {pages} {5679} (\bibinfo {year} {2019})},\ \Eprint {http://arxiv.org/abs/1906.08419} {arXiv:1906.08419 [astro-ph.GA]} \BibitemShut {NoStop}%
\bibitem [{\citenamefont {{Sofue}}(2020)}]{Sofue2020}%
  \BibitemOpen
  \bibfield  {author} {\bibinfo {author} {\bibfnamefont {Y.}~\bibnamefont {{Sofue}}},\ }\href {\doibase 10.3390/galaxies8020037} {\bibfield  {journal} {\bibinfo  {journal} {Galaxies}\ }\textbf {\bibinfo {volume} {8}},\ \bibinfo {eid} {37} (\bibinfo {year} {2020})},\ \Eprint {http://arxiv.org/abs/2004.11688} {arXiv:2004.11688 [astro-ph.GA]} \BibitemShut {NoStop}%
\end{thebibliography}%

\section{Supplemental Materials}

\subsection{Theoretical lower limits on the mixing angle}

While the upper limits on the sterile-active mixing $\theta$ inferred from the analysis of the $X$-ray Galactic spectrum depend only on the assumption that the sterile neutrinos form the dominant component of Galactic dark matter, there are lower limits on the same mixing related to a particular dark matter production mechanism operating in the early Universe and/or inherent in particular extensions of the SM where this mechanism is successfully implemented. 

To illustrate the relevance of the obtained limits, we consider the production of the sterile neutrinos in the primordial plasma with the lepton asymmetry, which allows for the {\it resonant amplification} of the active-to-sterile neutrino oscillations (similar to what we have for the solar neutrinos inside the Sun) and hence lower sterile-active mixing. The higher the asymmetry, the lower is the mixing angle required to produce the same amount of dark matter. The lepton asymmetry, which remained in the plasma after termination of the sterile neutrino production mechanism, typically gets redistributed between all three active neutrinos due to rapid neutrino oscillations at the plasma temperature of about 10\,MeV. Later, after neutrino decoupling from the primordial plasma, the {\it asymmetry in the electron neutrinos} impacts the neutron density, which determines the density of the helium nuclei produced in the primordial nucleosynthesis. Therefore, measurements of the primordial helium abundance constrain the asymmetry in the electron neutrinos and hence the asymmetry, which might help to produce the sterile neutrino dark matter. In turn, it imposes the lower limits on the sterile-active mixing angle required to produce the sufficient amount of the sterile neutrino dark matter particles.  

The lepton asymmetries can be characterized by the ratios $L_\alpha$ of the lepton number densities, $\alpha=e,\,\mu,\,\tau$, and entropy density $s$. After neutrino decoupling from the primordial plasma, the electron neutrino asymmetry can be approximately cast via the chemical potential $\mu_e$ and neutrino effective temperature $T_\nu$ as $n_{\nu_e}-n_{{\bar\nu}_e}=\mu_e T^2_\nu/6$. Present measurements on the helium abundance constrain the chemical potential as about $|\mu_e/T_\nu|<0.03$ at 95\% CL, see e.g. Ref.\cite{Escudero:2022okz}. It implies 
\[
L_e=\frac{n_{\nu_e}-n_{{\bar\nu}_e}}{s}=\frac{15}{4\pi^2 \cdot \frac{43}{11}}\cdot \frac{\mu_e}{T_\nu}\cdot \frac{4}{11}<1.06\times 10^{-3}\,
\]
where we take into account the difference between plasma and neutrino temperatures, $T_\nu=T\times (4/11)^{1/3}$, at this epoch. Before the neutrino decoupling from plasma, rapid oscillations between the active neutrinos redistribute any initial asymmetry between all the three lepton flavors in equal amount. Hence, if before the oscillations became rapid the initial lepton asymmetry $L_i$ was concentrated in only one flavor, say, in the muon sector, the above limit translates to 
\begin{equation}
\label{BBN-1}
L_{\mu,i}=3\, L_e<3.2\times 10^{-3}
\end{equation}

There are two numerical codes in literature, \texttt{sterile-dm}\,\cite{Venumadhav:2015pla} and \texttt{dmpheno}\,\cite{Ghiglieri:2015jua} based on different theoretical approaches, that calculate the sterile neutrino dark matter production in asymmetric plasma and give the required mixing angle and the asymmetry $L_{\mu,i}$ after the production terminates. In Fig.\,\ref{fig:final_c} the limit  "BBN-1" is obtained by running the numerical code \texttt{sterile-dm} (we thank D. Kalashnikov for sharing with us the numerical results) 
with final asymmetry in muon neutrinos 
obeying\,\eqref{BBN-1}. This line is very close to that obtained in Ref.\,\cite{Cherry:2017dwu}.  
The limit "BBN-2" is obtained upon proper rescaling 
of the line labeled as $L=2500$ in Fig.\,4 of Ref.\,\cite{Laine:2008pg}.  
Here we use the relation between the model parameters, 
$\sin^22\theta\times L\approx$\,const, observed for not small asymmetries $L$.

Finally, one can consider a more constrained situation, where the very asymmetry we need for the dark matter production is generated earlier by a mechanism implemented in a particular modification of the SM. The viable minimal example of this kind is provided by the $\nu$MSM model\,\cite{Asaka:2005an,Asaka:2005pn,Boyarsky:2009ix}, which is the SM supplemented with three sterile neutrinos. The lightest one is the dark matter candidate. Two others are heavier, of masses above 100\,MeV. They are responsible for the active neutrino getting masses via the type I seesaw mechanism. In the early Universe before the electroweak phase transition they produce the lepton asymmetry, which transfers to the baryonic sector by the electroweak sphalerons in the amount needed to explain the baryon asymmetry of the Universe. Later they come to equilibrium with plasma and then freeze-out and finally decay, producing much more lepton asymmetry needed for the dark matter sterile neutrino production. It is the self-contained phenomenologically viable model capable of explaining simultaneously neutrino masses, baryon asymmetry of the Universe and dark matter with only three Majorana fermions (sterile neutrinos) added to the SM. In this model, the maximal lepton asymmetry generated in the heavy sterile neutrino decays is much smaller than the limit \eqref{BBN-1}: the scan performed in Ref.\,\cite{Canetti:2012kh} revealed $L_\mu<1.24\times 10^{-4}$ and the corresponding limit on the mixing angle is shown in Fig.\,2 of the same paper. In Fig.\,\ref{fig:final_c} we depict this limit as "$\nu$MSM".   

Since the sterile neutrino dark matter particles are produced in oscillations of active neutrinos, which are in equilibrium in plasma, they inherit the momentum distribution with a typical value of order plasma temperature. For a particle in keV mass range, it implies a velocity of about $10^{-3}-10^{-4}$ at radiation-matter equality and hence some suppression of the cosmic structure formation at correspondingly small scales. The induced limits constrain the sterile neutrino mass from below but naturally depend on the dark matter distribution over momentum. The two neutrino codes \texttt{sterile-dm} and \texttt{dmpheno} give noticeably different distributions yielding different average momentum, which may be found in the cited above literature.

\subsection{Observation and data processing}

Figure~\ref{fig:poly} shows the 11-year light curve of FPMA and FPMB based on their 3$-$10~keV detector count rate in each observation. One can notice characteristic spikes, mainly caused by the observations in the Galactic plane, where GRXE contributes to the detector count rate, solar activity or possible imperfection of the detector cleaning procedure. The long-term variation related to slow change of the radiation environment is also clearly seen. To take the latter into account, we model the light-curve by the cubic polynomial form using an outlier-resistant fitting procedure. It allows us to apply a continuously changing threshold for excluding observations with a high detector count rate. In our analysis, we exclude all ObsIDs with detector count rates greater than a factor of 2 from the polynomial fit (shown as a dashed curved line). Such detector light curve filtering reduces the total exposure of $|b|>3^{\circ}$ and $|b|<3^{\circ}$ observations by 5\% and 47\%, respectively. The stronger cut at low Galactic latitudes is explained by the fact that detector count rate contains flux from the GRXE.

\begin{center} 
{\it  The concept of stray light data analysis} 
\end{center}

The open structure of the \nustar's mast allows photons with off-axis angles ${\sim}1-3^{\circ}$ to strike the detectors without being focused by the mirror optics. This stray light (SL) contamination led to enhanced background and, consequently, suppressed sensitivity for point-like and extended X-ray sources. However, SL background contains information about X-ray surface brightness of the sky used in many astrophysical studies \citep[see][for a review]{2023JATIS...9d8001M}. 

\begin{figure}[htb]
    \centering
    \includegraphics[width=0.9\columnwidth]{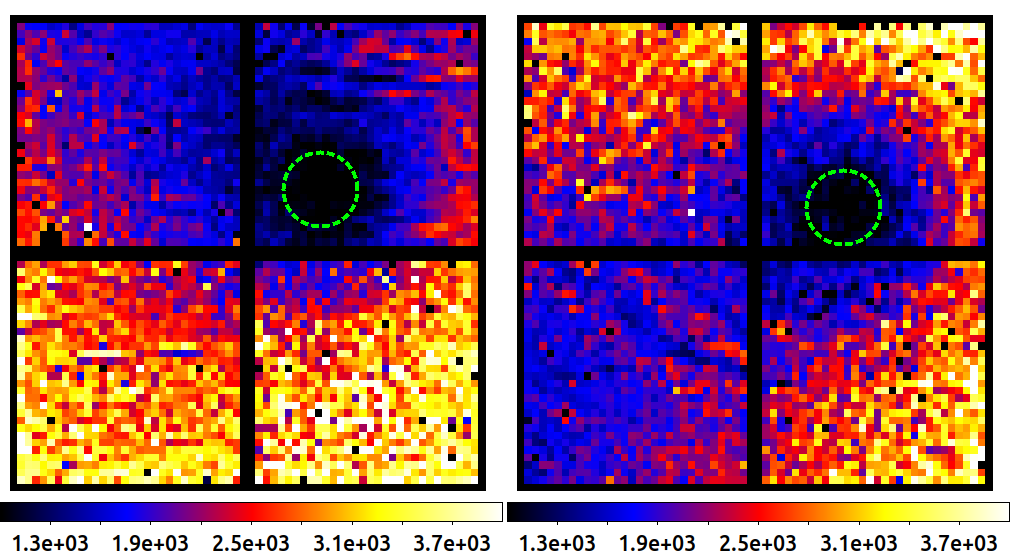}
    \caption{Stacked FPMA (left) and FPMB (right) images in total counts detected in physical detector pixels, based on all cleaned {\nustar} observations at $|b|>3^{\circ}$ with a total exposure of 117~Ms/FPM in the 3–20 keV energy band. The dashed circle shows the approximate position of the optical axis.}
    \label{fig:stray}
\end{figure}

\begin{figure}
  \includegraphics[width=0.99\columnwidth]{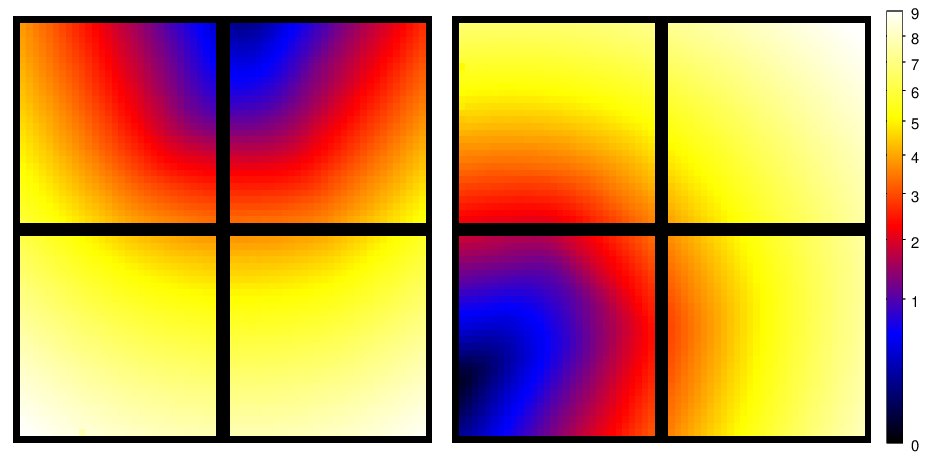}
\caption
{ \label{fig:deg2} 
Image of the \nustar\ FPMA (left) and FPMB (right) in physical detector pixels, showing the open portion of the sky for each detector pixel in squared degrees. The image is adopted from \cite{Krivonos2021}.}
\end{figure}

{\nustar} SL background generates a specific spatial pattern on the detectors, which is a result of the convolution of the sky and collimated angular response function of the {\nustar}, modified by the optical bench structure \cite{Wik:2014boa}. The spatial pattern is different for the two FPMs since they have different views on the optical bench. Fig.~\ref{fig:stray} shows stacked FPMA and FPMB images based on all cleaned {\nustar} observations at $|b|>3^{\circ}$ with a total exposure of 117~Ms/FPM in the 3–20 keV energy band. Both FPMA and FPMB images demonstrate strong spatial gradient. The suppression at the approximate position of the optical axis (marked by the dashed circle) is caused by removed target X-ray sources. The SL spatial pattern is caused by the fact that each detector pixel is open to the sky at different solid angles, ranging from ${\sim}0.2$ to ${\sim}9$ deg$^2$, as shown in Fig.~\ref{fig:deg2}. 

The effective solid angle of the {\nustar} SL aperture $\Omega_{\rm SL}$, averaged over the full FOV is about $4.5$~deg$^{2}$. The corresponding detector effective area for SL photons $A_{\rm SL}$ is limited by the physical ${\sim}13$~cm$^{2}$ area of FPMA and FPMB detector arrays. The product of the FOV multiplied by the (average) effective area is usually referred to as a ``grasp'', which characterizes the light gathering power of the X-ray telescope, especially important for X-ray surveys and studies of the X-ray surface brightness of the sky (like the current one). The SL grasp of {\nustar} $\Omega_{\rm SL}\times A_{\rm SL}$ can ideally reach the value of ${\sim}58$~deg$^2$~cm$^2$ for each module, however, after bad pixel removal, the usable detector area and effective solid angle for a typical ``clean'' observation decrease, respectively, to $A_{\rm SL}{\sim}10$~cm$^2$ and $\Omega_{\rm SL}{\sim}4$~deg$^2$ \cite{Perez:2019bgo}, which makes the effective SL grasp at the level of ${\sim}40$~deg$^2$~cm$^2$. Nevertheless, this value is significantly larger than the grasp of the photons focused by the {\nustar}'s mirror optics ${\approx}8$~deg$^2$~cm$^2$ per FPM \citep[see e.g.][]{Roach:2022lgo}. Thus, {\nustar} SL aperture can be effectively used for studies of diffuse X-rays on the scale of several degrees.

The {\nustar} detector count rate contains essentially two basic components: 
\begin{itemize}
    \item Spatially flat detector background: The count rate in a given pixel is proportional to a nearly flat instrumental background, which includes internal emission lines and continuum;
    \item Spatially variable aperture component: The count rate is
proportional to the open sky solid angle.
\end{itemize}
The concept of the method is to separate out the flat and aperture components\footnote{Note that, by design of this model, we neglect the focused CXB and contribution of undetected point sources.}. The method implies maximization of a likelihood function constructed as follows. The expected total counts $N_{\rm pix}$ observed in the $i^{\rm th}$ detector pixel, in the $E_\gamma$ energy interval, during exposure time $T$ is given by: 
\begin{equation}
    N_{\text{pix},i}(E_\gamma) = (C_\text{int} M_\text{int} Q + C_\text{apt} R_\text{pix} \mathcal{E}_\text{tot} A \Omega)_i T
\end{equation}
where $C_\text{int,i}$ is the internal background count rate, $M_\text{int,i}$ describes detector uniformity \citep[see][for details]{Krivonos2021}, $Q$ is the relative time-dependent correction factor tracing the long-term radiation environment variation (see above and Fig.~\ref{fig:poly}), $C_\text{apt,i}$ is the aperture flux per solid angle, $R_\text{pix}$ is the pixel response matrix stored in the \nustar\ CALDB, $\mathcal{E}_\text{tot}=\mathcal{E}_\text{det}\mathcal{E}_\text{Be}$ is the energy-dependent efficiency of the inactive detector surface layer and beryllium entrance window, $A$ is the area of each detector pixel $(0.6~\text{mm})^2$, $\Omega_i$ is the open sky solid angle as seen by each pixel in deg$^2$, $T$ is exposure time. To construct the spectrum of $C_\text{apt}$ we defined 100 energy bands $E_\gamma$ logarithmically spaced between 3 and 20 keV.

We use a log-likelihood function to derive the maximum likelihood estimator of the parameters $C_{\rm int}$ and $C_{\rm apt}$:
\begin{equation}
\label{eq:likelihood}
L = -2 \sum_{\rm i}  N_{\rm i}*\log N_{\rm pix, i} - N_{\rm pix,i} - \log N_{\rm i}! 
%-2\Sum( ima_1d*alog(pcxb)-pcxb-alog(factorial(ima_1d)) )
\end{equation}
where $i$ runs over the array of pixels (excluding bad pixels) for all observations, $N_{\rm i}$ is the observed number of counts in the $i^{\rm th}$ pixel. The calculation of the aperture FOV parameter $\Omega$ for each detector pixel (Fig.~\ref{fig:deg2}) is based on the known structure of the telescope and done with the {\sc nuskybgd} code \cite{Wik:2014boa}.

\begin{center} 
{\it  North Ecliptic Pole (NEP)} 
\end{center}

\begin{figure}[tb]
    \centering
    \includegraphics[width=\columnwidth]{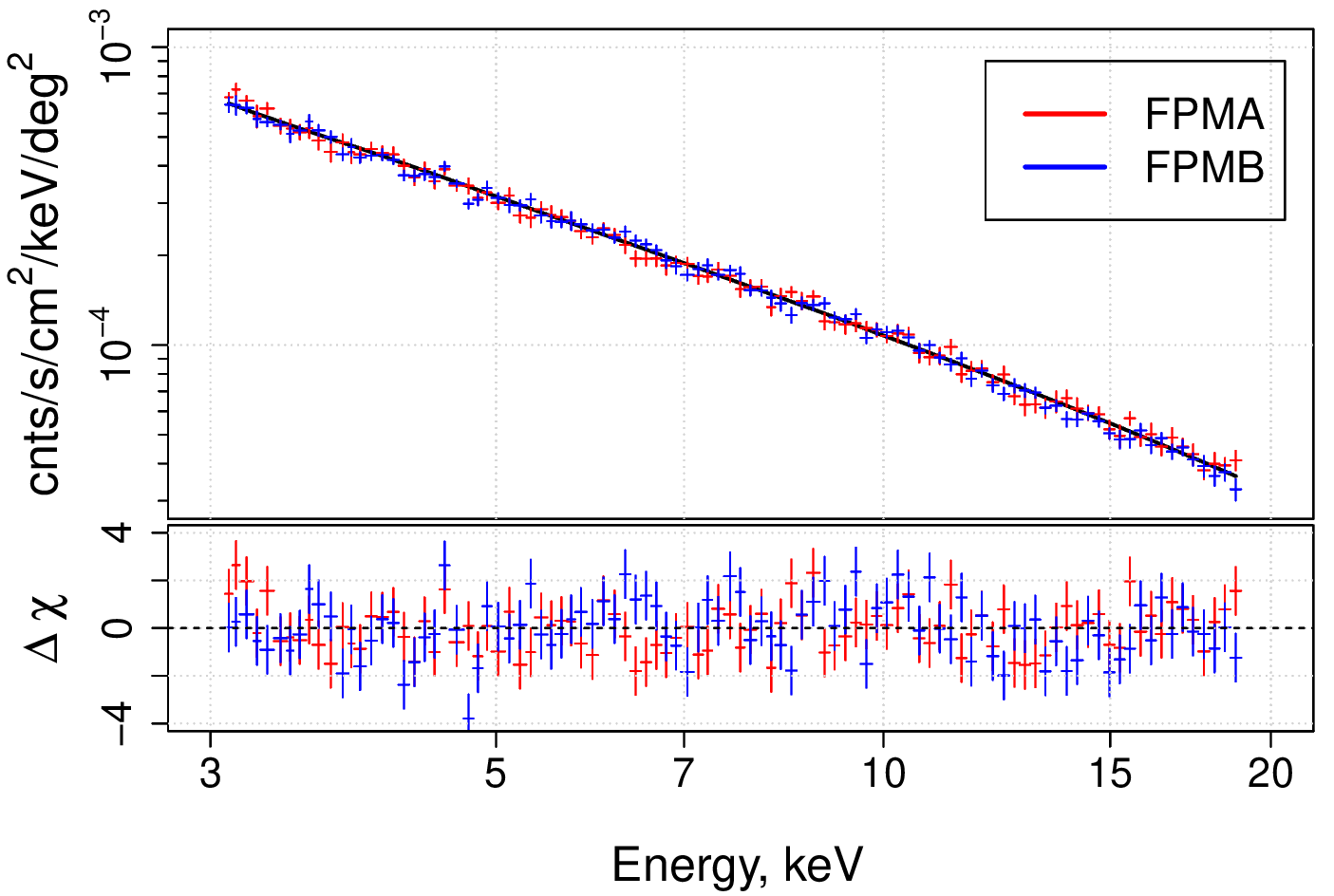}
    \caption{The \nustar\ FPMA (red) and FPMB (blue) spectra of the aperture sky background (mainly dominated by CXB), measured in NEP-TDF field.}
    \label{fig:nep}
\end{figure}

To validate the current analysis, we applied our method to the data of the ongoing \nustar\ North Ecliptic Pole (NEP) extragalactic survey \cite{Zhao:2024klp} of the JWST NEP Time-Domain Field \citep[TDF,][]{Windhorst:2022rrv}. The \nustar\ NEP-TDF survey is a multi-year program started in 2019.  The current \nustar\ data set of NEP-TDF observation contains 35 observations with total exposure time of 2.7~Ms/FPM. This extragalactic field is free from bright sources, which allows us to directly extract the aperture background component dominated by CXB. Figure\,\ref{fig:nep} shows FPMA and FPMB spectra of the aperture background. Hereafter we ignore energy bins above 19 keV due to increased aperture flux caused by the so called Absorbed Stray Light \citep[ASL,][]{2017JATIS...3d4003M,Rossland:2023vfc,Weng:2024rnt}. Similar to \cite{Krivonos2021}, the spectrum is well approximated by two components: a low-energy ($E<5$~keV) power-law component with a steep photon index $\Gamma_{\rm sol}\simeq5-7$ attributed to the solar emission reflected from the back of the aperture stops, and a power-law with $\Gamma_{\rm cxb}=1.29$ modified by a high energy cutoff with $E_{\rm cut}=41.13$~keV, representing the canonical CXB spectrum parameterized by \cite{1999ApJ...520..124G}. The fitting procedure with fixed $\Gamma_{\rm sol}=4$, $\Gamma_{\rm cxb}$, $E_{\rm cut}$ and free normalization parameters yields fit statistics $\chi^{2}_{\rm r}$/dof =$259.52/191=1.36$. We noticed that the fit quality can be improved to $\chi^{2}_{\rm r}$/dof =$229.21/190=1.21$ by releasing the high-energy cutoff parameter, estimated as $E_{\rm cut}=31.72_{-2.04}^{+2.34}$~keV. The inferred normalizations $F_{\rm 3-20\,keV}^{\rm FPMA}=(2.89\pm0.03)\times 10^{-11}$~\ergscm\ and $F_{\rm 3-20\,keV}^{\rm FPMB}=(2.87\pm0.03)\times 10^{-11}$~\ergscm\ are consistent with measurements\,\cite{Krivonos2021}, taking statistical and systematic uncertainties into account. We should also stress here that FPMA and FPMB spectra are consistent with each other, which validates the whole method, because \nustar\ FPMs observe practically the same sky region. In our analysis we combine FPMA and FPMB data sets, which corresponds to linking normalisation of the aperture background seen by FPMs.

\begin{center} 
{\it Galactic sky $|b|<3^{\circ}$}  
\end{center} 

Studying the CXB emission with \nustar\ stray light data, Ref.\,\cite{Rossland:2023vfc} noticed the appearance of a line-like component around the Fe complex at 6.7\,keV, visible up to Galactic latitudes $|b|\approx30^\circ$, which pointed out a possible contribution from the GRXE. 
In the current analysis, we do not see any significant excess around the position of the Fe complex in the {\acxb} spectrum at $|b|>3^{\circ}$. This can be explained by the fact that in the current analysis we excluded observations with a high detector count rate, mainly caused by the contribution from the GRXE. To demonstrate the impact of GRXE on the \nustar\ stray light spectrum, we selected all \nustar\ observations at $|b|<3^{\circ}$  and,  for consistency,  applied the same detector curve filtering. This data set, hereafter referred to as \textit{GAL}, gathered 26~Ms of the combined FPMA+FPMB exposure (236 FPMA and 266 FPMB observations). 

\begin{figure}[th]
    \centering
    \includegraphics[width=\columnwidth]{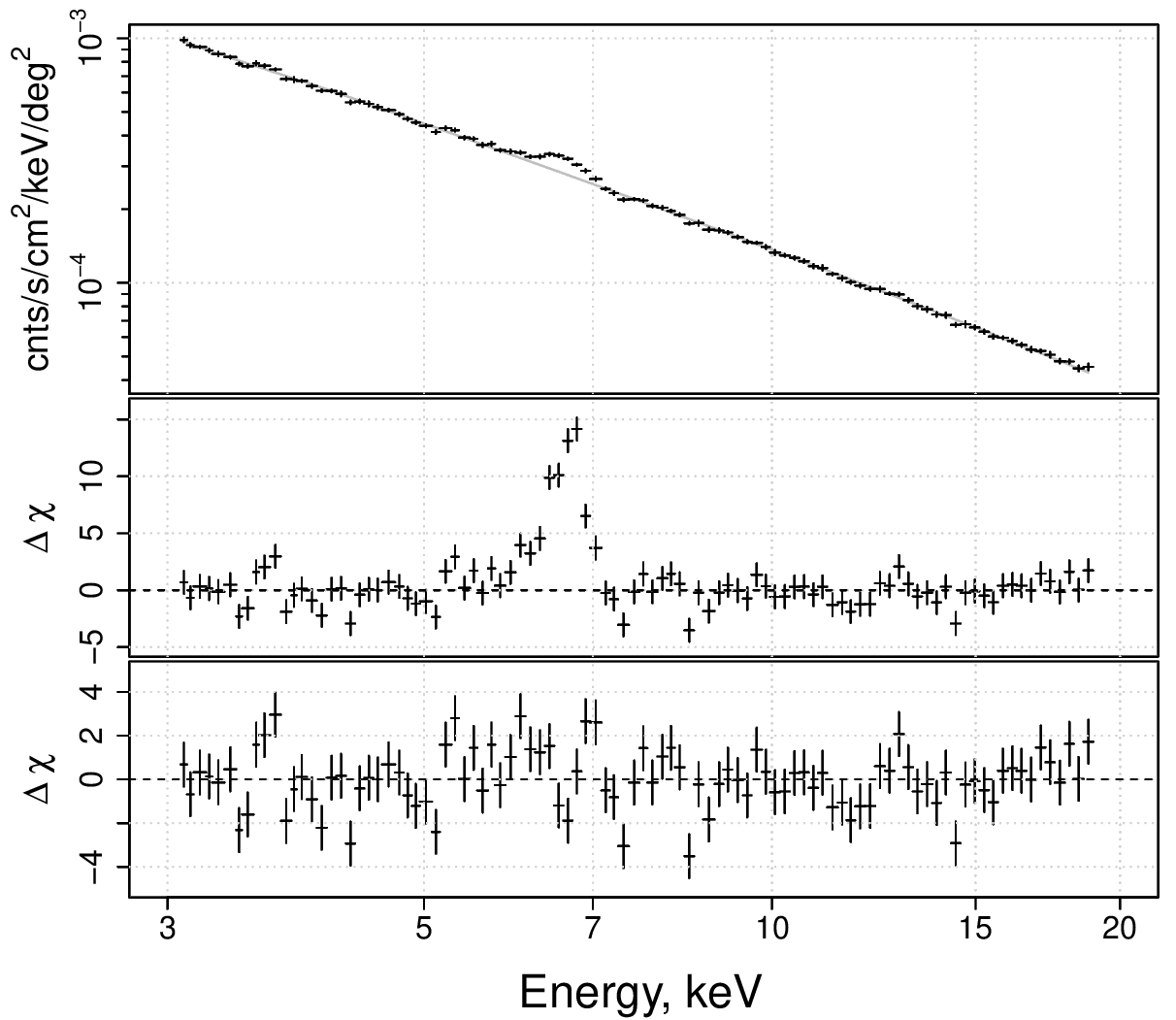}
    \caption{The \nustar\ combined FPMA and FPMB stray-light spectrum, measured at $|b|<3^{\circ}$ (black data points, upper panel). Spectral model \texttt{powerlaw + cutoffpl} without Gaussian line at 6.7~keV is shown by the gray line. The residuals of this model are demonstrated on the middle panel. The bottom panel shows the residuals with Gaussian line added.}
    \label{fig:gal3}
\end{figure}   

Figure~\ref{fig:gal3} shows \textit{GAL} stray light spectrum. We first fitted it with a sum of a power-law with $\Gamma=4$, representing solar component and phenomenological power-law with a high-energy cut-off (\texttt{cutoffpl} in XSPEC notation). The spectral fit is worse ($\chi^{2}_{\rm r}$/dof =$722.30/92=7.85$), mainly because of strong line-like excess around 6.7~keV, clearly visible in the residuals (Fig.~\ref{fig:gal3}, middle panel). The GRXE contains lines of ionized iron at energies ${\sim}6.7$-$6.9$~keV  with a very large equivalent width (${\sim}1$~keV) \cite{Revnivtsev:2006gf}. As a simple representation of the complex of emission lines at 6.4$-$6.9\,keV, we added a Gaussian line, which significantly improved the quality of the fit ($\chi^{2}_{\rm r}$/dof =$144.91/90=1.61$). The \texttt{cutoffpl} component is described with $\Gamma=1.61\pm0.07$, poorly constrained high-energy cut-off $E_{\rm cut}=65_{-18}^{+37}$~keV, and relatively high flux $F_{\rm 3-20\,keV}=(3.84\pm0.04)\times 10^{-11}$~\ergscm\ (note that \textit{GAL} spectrum contains a mixture of CXB and GRXE). The line component is described by a position at $E=6.56\pm0.02$~keV and width $\sigma=0.22\pm0.04$~keV. The 90\% confidence interval of the equivalent width $0.12 - 0.15$\,keV is less than expected, however, we would avoid astrophysical interpretation because we excluded a large fraction of data with a strong contribution of GRXE. The detailed analysis of the GRXE component in the \nustar\ stray light background will be presented elsewhere.

% initial selection bmax=3, area_max=0.4, cstat=1.4
%fix_gal_fpmA total exposure        26090.525 ks, total ObsID 507
%fix_gal_fpmB total exposure        23628.109 ks, total ObsID 459

% after poly_cut=2.0
% Total used ObsIDs 502 from 965
% Total used exposure, 26486.87 ks

\subsection{Calculation of the signal flux}

\begin{figure}[!htb]
	\centering
	\includegraphics[width=0.9\linewidth]{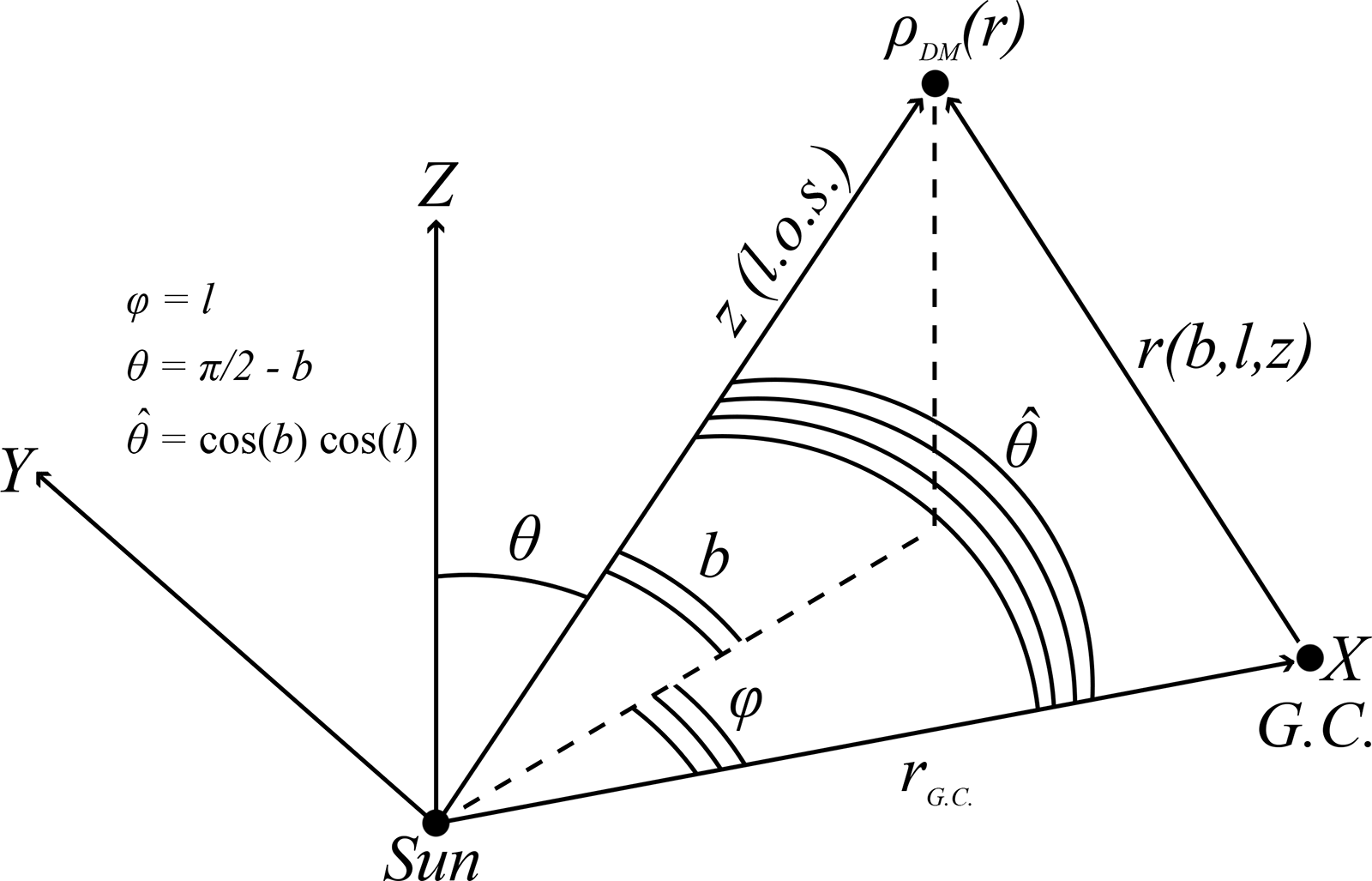}
	\caption{The relationship between galactic and spherical coordinates to calculate $\mathcal{D}_{DM}$ for a given object in the sky. %Calculations for an arbitrary solid angle shape and for an arbitrary object in the sky are conveniently performed in the galactic coordinate system relative to the location of the object from the Milky Way.
 }
	\label{fig:gal_sph_img}
\end{figure}

The relationship between the spherical coordinate system and the galactic coordinate system, as well as the angle $\hat{\theta}$ indicating the direction from the center of the Milky Way to a given object (the angle between $z$ and the $X$ axis, see Fig~\ref{fig:gal_sph_img}), is given by
\begin{equation}
 \text{d} \Omega = \sin\left(\frac{\pi}{2} - b\right)\text{d}b\text{d}l\,,\;\;\; \phi = l\,,\;\;\; \theta = \frac{\pi}{2} - b\,.
\end{equation}

\begin{figure*}[!htb]
	\centering
	\subfigure{\includegraphics[width=0.49\linewidth]{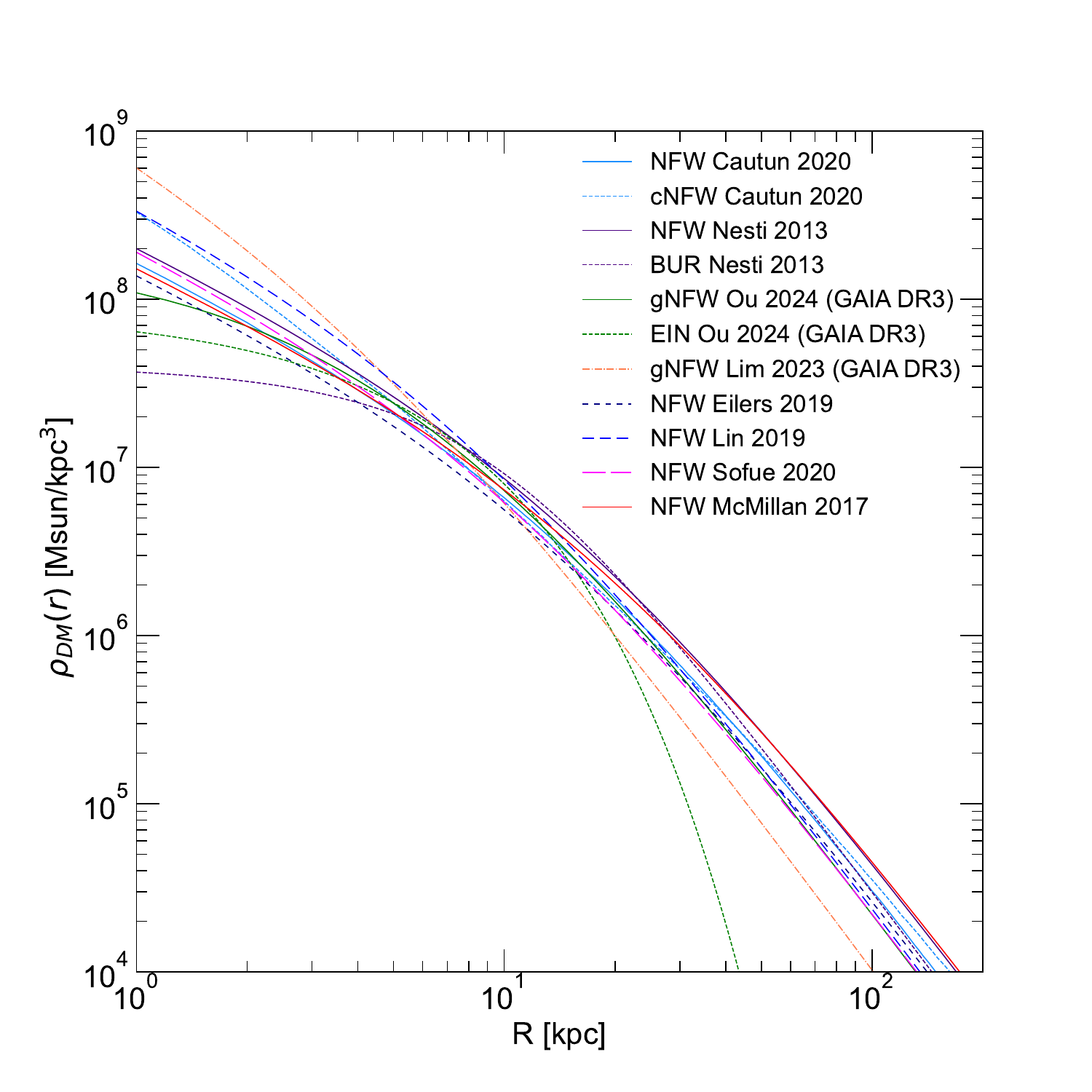}}
% \end{figure}
\hspace{-1cm}
% \begin{figure}[!htb]
	\centering
	\subfigure{\includegraphics[width=0.49\linewidth]{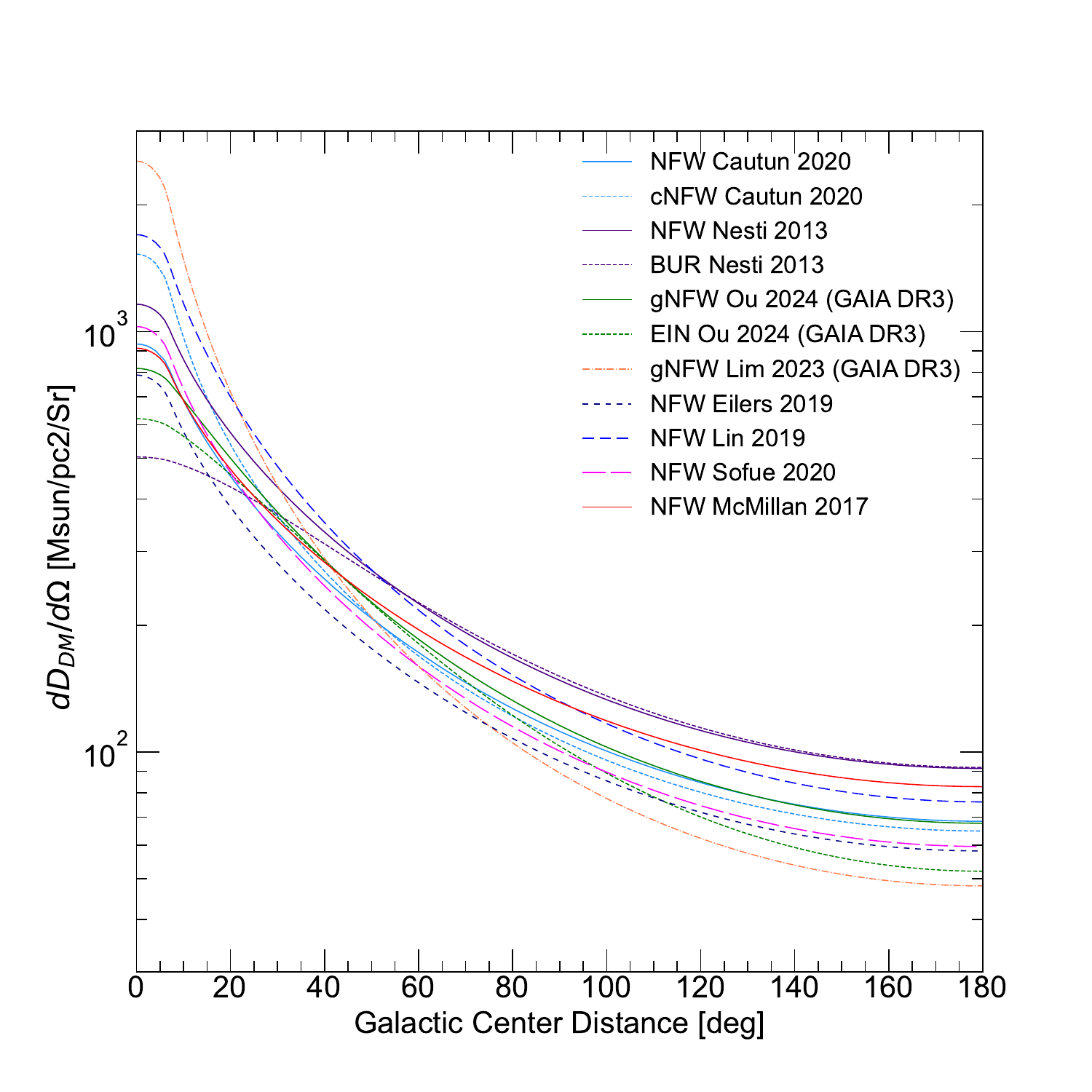}}
	\caption{\textbf{Left:} The dark matter density profiles presented in Table~\ref{tab:diff_column_densities}. \textbf{Right:} The differential column density profiles calculated for each dark matter profiles.}
	\label{fig:dm_profiles_and_column_densities}
\end{figure*}

\begin{center}    
{\it Signal dependence on the DM distribution}
\end{center}
We perform our analysis on a wide range of dark matter density profiles. Figures~\ref{fig:dm_profiles_and_column_densities} present the dark matter density profiles and differential column densities for each of the profiles used in our analysis. We fix dark matter density at the small distances from the galaxy center as $\rho_{DM}(r < 1 \text{kpc}) = \rho_{DM}(r = 1 \text{kpc})$. It does not noticeably affect our analysis since we consider only directions at $|b|>3^\circ$. In Table~\ref{tab:diff_column_densities} we present the stacked differential column density values for each of the profiles that we use to calculate the $X$-ray intensity from sterile neutrino decays. The field of view of each observation is quite small and most directions are far from the galactic plane, so the approximation $\mathcal{D}_{DM, i} \approx (d\mathcal{D}_{DM, i} / d\Omega) \Delta \Omega_i$ works to a very good accuracy, though we do not use it in our analysis.

%
%
% \begin{figure}[htb]
%   \centering
%   \subfigure{\includegraphics[scale=0.35]{figures/dark_matter_profiles.pdf}}
%   % \vspace*{-5.0cm}
%   \subfigure{\includegraphics[scale=0.35]{figures/column_densities_Msunpc2Sr.pdf}}
%   \caption{\textbf{Left:} The dark matter density profiles presented in Table~\ref{tab:diff_column_densities}. \textbf{Right:} The differential column density profiles calculated for each of the profiles presented in the left panel.}
%   \label{fig:dm_cd_profiles}
% \end{figure}
\begin{table}[!htb]
    \centering
    \caption{The values of stacked differential column density for each type of the dark matter profile. }
    \begin{tabular}{ccrc}
    \hline
    \hline

    $\langle \text{d} \mathcal{D}_{DM} / \text{d} \Omega \rangle$ & $\langle \text{d} \mathcal{D}_{DM} / \text{d} \Omega \rangle $  & & \\
    
    [$\text{M}_{\odot} \hspace{1pt} \text{pc}^{-2} \hspace{1pt} \text{Sr}^{-1}$] & [$\text{GeV} \hspace{1pt} \text{cm}^{-3} \hspace{1pt} \text{kpc} \hspace{1pt} \text{Sr}^{-1}$] & Profile & Refs \\
    \hline
    \hline
    140.2 & 5.32  &  NFW  &  \cite{Cautun:2019eaf}\\
    145.7 & 5.53  & cNFW  &  \cite{Cautun:2019eaf}\\
    183.1 & 6.95  &  NFW  &  \cite{Nesti2013}\\
    171.5 & 6.51  &  BUR  &  \cite{Nesti2013}\\
    146.6 & 5.57 & gNFW  &  \cite{Ou2024}\\
    131.8 & 5.00  &  EIN  &  \cite{Ou2024}\\
    148.7 & 5.65  & gNFW  &  \cite{Lim2023}\\
    118.9 & 4.52  &  NFW  &  \cite{Eilers2019}\\
    181.3 & 6.88  &  NFW  &  \cite{Lin2019}\\
    131.2 & 4.98 &  NFW  &  \cite{Sofue2020}\\
    158.6 & 6.02  &  NFW  &  \cite{McMillan2017}\\
    \hline
    \hline
    \end{tabular}
    \label{tab:diff_column_densities}
\end{table}

\begin{center}
{\it Spectral model and data analysis}
\end{center}

In Tab.\,\ref{tab:bf_pars} we present the best fit values for the base model parameters used in our analysis.

\begin{table}
    \centering
    \caption{The values of the best fit parameters used in our base model for the {\acxb} spectrum (90\% confidence).}
    \begin{tabular}{llll}
    \hline
    \hline
         Model & Parameter & Value & Frozen \\
    \hline
    \hline
         \texttt{powerlaw} & $\Gamma_{\text{sol}}$ & 4      & True\\
         %\texttt{powerlaw} & $N_{\text{sol}}$      & 9.8029$\times10^{-3}$ $\pm$ 2.9$\times10^{-4}$  & False\\
         \texttt{powerlaw} & $N_{\text{sol}}^\text{a}$      & $(9.8\pm0.3)\times10^{-3}$   & False\\
         \texttt{cflux}    & $E_{\text{min}}$      & 3~keV      & True\\
         \texttt{cflux}    & $E_{\text{max}}$      & 20~keV      & True\\
         %\texttt{cflux}    & $lg_{10}$Flux              &-10.5196 $\pm$ 4.8$\times10^{-4}$       & False\\
         \texttt{cflux}    & Flux$^\text{b}$               & $(3.023\pm0.006)$       & False\\
         \texttt{powerlaw} & $\Gamma_{\rm cxb}$ & 1.29      & True\\
         %\texttt{powerlaw} & $N_{\rm cxb}$      & 2.39933$\times 10^{-3}$  & True\\
         \texttt{highecut} & $E_{\rm cut}^{\rm cxb}$      & $10^{-4}$~keV  & True\\
         %\texttt{highecut} & $E_{\text{fold}}$     & 34.8765 $\pm$ 0.34      & False\\
         \texttt{highecut} & $E_{\text{fold}}$     &    $34.9\pm0.6$\,keV   & False\\
    \hline
    \hline
     Test statistic: & $\chi^2/\text{dof} = 1.38$, & p = $8.27 \times 10^{-3}$  \\
    \hline
    \hline
    \end{tabular}
    \label{tab:bf_pars}
    \begin{flushleft}
    $^\text{a}$ photons keV$^{-1}$~s$^{-1}$~cm$^{-2}$~deg$^{-2}$ at 1~keV.\\
    $^\text{b}$ $\times 10^{-11}$~\ergscm    
    \end{flushleft}
    
\end{table}
% Similar to NEP case described in SuMs, we performed fitting procedure with fixed $\Gamma_{\rm sol}=4$, $\Gamma_{\rm cxb}$, $E_{\rm cut}$ and allowing normalization parameters and high-energy cutoff to vary. The fit is characterised by $\chi^{2}_{\rm r}$/dof = $130.00/94$ = $1.38$. The high-energy cutoff was estimated at $E_{\rm cut}=34.9\pm0.6$\,keV. The CXB normalization was measured as $F_{\rm 3-20\,keV}=(3.023\pm0.006)\times 10^{-11}$~\ergscm, which is somewhat higher than measurements \cite{Krivonos2021}, but still consistent taking overall systematic uncertainty into account.

\end{document}